\documentclass{emulateapj}

\newcommand{\sqcm}{${\rm cm^{-2}}$}

\newcommand{\mum}{${\rm \mu m}$}
  
\newcommand{\waven}{${\rm cm^{-1}}$}

\slugcomment{}

\shorttitle{Spitzer Space Telescope Survey of Interstellar Ices}
\shortauthors{Boogert et al.}

\received{23/Sep/2007}
\accepted{05/Jan/2008}

\begin{document}

\title{The c2d Spitzer Spectroscopic Survey of Ices Around Low-Mass
  Young Stellar Objects: I. H$_2$O and the 5-8 \mum\
  Bands\altaffilmark{1,2}}

\author{A. C. A. Boogert\altaffilmark{3,4,5},
        K. M. Pontoppidan\altaffilmark{6,7},
        C. Knez\altaffilmark{8},
        F. Lahuis\altaffilmark{9,10},
        J. Kessler-Silacci\altaffilmark{11},
        E. F. van Dishoeck\altaffilmark{9},
        G. A. Blake\altaffilmark{6},
	J.-C. Augereau\altaffilmark{12},
        S. E. Bisschop\altaffilmark{9},
        S. Bottinelli\altaffilmark{9},
        T. Y. Brooke\altaffilmark{13},
        J. Brown\altaffilmark{3},
        A. Crapsi\altaffilmark{9},
        N. J. Evans, II\altaffilmark{11},
        H. J. Fraser\altaffilmark{14},
        V. Geers\altaffilmark{9},
        T. L. Huard\altaffilmark{15},
	J. K. J{\o}rgensen\altaffilmark{15},
        K. \"Oberg\altaffilmark{9},
        L. E. Allen\altaffilmark{15},
        P. M. Harvey\altaffilmark{11},
        D. W. Koerner\altaffilmark{16},
        L. G. Mundy\altaffilmark{8},
        D. L. Padgett\altaffilmark{13},
        A. I. Sargent\altaffilmark{3},
	K. R. Stapelfeldt\altaffilmark{17}}

\altaffiltext{1}{Some of the data presented herein were obtained at
    the W.M. Keck Observatory, which is operated as a scientific
    partnership among the California Institute of Technology, the
    University of California and the National Aeronautics and Space
    Administration. The Observatory was made possible by the generous
    financial support of the W.M. Keck Foundation.}
\altaffiltext{2}{The
    VLT/ISAAC spectra were obtained at the European Southern
    Observatory, Paranal, Chile, within the observing programs
    164.I-0605, 69.C-0441, and 272.C-5008}
\altaffiltext{3}{Division of PMA, Mail Code 105-24, California
  Institute of Technology, Pasadena, CA 91125, USA}
\altaffiltext{4}{AURA/NOAO-South, Gemini Science Center, Casilla 603,
  La Serena, Chile}
\altaffiltext{5}{current address: IPAC, NASA Herschel Science Center,
  Mail Code 100-22, California Institute of Technology, Pasadena, CA
  91125, USA (email: aboogert@ipac.caltech.edu)}
\altaffiltext{6}{Division of GPS, Mail Code 150-21, California
  Institute of Technology, Pasadena, CA 91125, USA}
\altaffiltext{7}{Hubble Fellow}
\altaffiltext{8}{Department of Astronomy, University of Maryland,
  College Park, MD 20742, USA}
\altaffiltext{9}{Leiden Observatory, PO Box 9513, 2300 RA Leiden, the
  Netherlands}
\altaffiltext{10}{SRON, PO Box 800, 9700 AV Groningen, the Netherlands}
\altaffiltext{11}{Department of Astronomy, University of Texas at
  Austin, 1 University Station C1400, Austin, TX 78712-0259, USA}
\altaffiltext{12}{Laboratoire d'Astrophysique de Grenoble, CNRS,
  Universit\'e Joseph-Fourier, UMR 5571, Grenoble, France}
\altaffiltext{13}{Spitzer Science Center, California Institute of
  Technology, CA 91125, USA}
\altaffiltext{14}{Department of Physics, Scottish Universities Physics
  Alliance (SUPA), University of Strathclyde, John Anderson Building,
  107 Rottenrow East, Glasgow G4 ONG, Scotland, UK}
\altaffiltext{15}{Smithsonian Astrophysical Observatory, 60 Garden
  Street, MS42, Cambridge, MA 02138, USA}
\altaffiltext{16}{Department of Physics and Astronomy, Northern Arizona
  University, Box 6010, Flagstaff, AZ 86011-6010, USA}
\altaffiltext{17}{Jet Propulsion Laboratory, MS 183-900, California
  Institute of Technology, 4800 Oak Grove Drive, Pasadena, CA 91109, USA}

\begin{abstract}

  With the goal to study the physical and chemical evolution of ices
  in solar-mass systems, a spectral survey is conducted of a sample of
  41 low luminosity YSOs ($L\sim 0.1-10~{\rm L}_{\odot}$) using 5--38
  \mum\ Spitzer Space Telescope and 3--4 \mum\ ground-based spectra.
  The sample is complemented with previously published Spitzer spectra
  of background stars and with ISO spectra of well studied massive
  YSOs ($L\sim 10^5~{\rm L}_{\odot}$).  This paper focuses on the
  origin of the prominent absorption features in the 5-8 \mum\
  spectral region.  The long-known 6.0 and 6.85 \mum\ bands are
  detected toward all sources, with the Class 0-type low mass YSOs
  showing the deepest bands ever observed.  In almost all sources the
  6.0 \mum\ band is deeper, by up to a factor of 3, than expected from
  the bending mode of pure solid H$_2$O, based on the optical depths
  of the 3.0 \mum\ stretching and 13 \mum\ libration modes.  The depth
  and shape variations of the remaining 5--7 \mum\ absorption indicate
  that it consists of 5 independent components, which, by comparison
  to laboratory studies, must be from at least 8 different carriers.
  Together with information from the 3-4 \mum\ spectra and the
  additionally detected weak 7.25, 7.40, 9.0, and 9.7 \mum\ features
  it is argued that overlapping bands of simple species are
  responsible for much of the absorption in the 5-7 \mum\ region, at
  abundances of 1-30\% for CH$_3$OH, 3-8\% for NH$_3$, 1-5\% for
  HCOOH, $\sim$6\% for H$_2$CO, and $\sim$0.3\% for HCOO$^-$ with
  respect to solid H$_2$O.  The 6.85 \mum\ band likely consists of one
  or two carriers, of which one is less volatile than H$_2$O because
  its abundance relative to H$_2$O is enhanced at lower
  H$_2$O/$\tau_{9.7}$ ratios. It does not survive in the diffuse
  interstellar medium (ISM), however.  The similarity of the 6.85
  \mum\ bands for YSOs and background stars indicates that its
  carrier(s) must be formed early in the molecular cloud evolution.
  If an NH$_4^+$ salt is the carrier its abundance with respect to
  solid H$_2$O is typically 7\%, and low temperature acid-base
  chemistry or cosmic ray induced reactions must have been involved in
  its formation.  Possible origins are discussed for the carrier of an
  enigmatic, very broad absorption between 5 and 8 \mum.  It shows
  large depth variations toward both low- and high-mass YSOs.  Weak
  evidence is found that it correlates with temperature tracers.
  Finally, all the phenomena observed for ices toward massive YSOs are
  also observed toward low mass YSOs, indicating that processing of
  the ices by internal ultraviolet radiation fields is a minor factor
  in the early chemical evolution of the ices.

\end{abstract}

\keywords{infrared: ISM --- ISM: molecules --- ISM: abundances ---
  stars: formation --- infrared: stars--- astrochemistry}

\section{Introduction}~\label{sec:intro}

The infrared spectra of protostars and obscured background stars show
prominent absorption features at 3.0, 4.25, 4.7, 6.0, 6.85, 9.7, and
15 \mum\ along with a suite of weaker features (e.g. \citealt{dhe96},
\citealt{whi96}, \citealt{ger99}, \citealt{sch99}, \citealt{gib00},
\citealt{kea01b}, \citealt{pon03a}, \citealt{gib04}, \citealt{kne05},
\citealt{whi07}; see \citealt{boo04} for a complete list of features).
These are attributed to absorption in the vibrational modes of
molecules in ices, except for the 9.7 \mum\ band which is mostly due
to silicates. At the low temperatures ($T\leq$20 K) of dense clouds
and circum-protostellar environments atoms and molecules freeze out
rapidly on dust grains.  Grain surface chemistry efficiently forms
new, simple species, such as H$_2$O and H$_2$CO (e.g.
\citealt{tie82}).  Complex species (e.g.  polyoxymethylene [`POM',
-(CH$_2$-O)$_{\rm n}$-], and hexamethylenetetramine ['HMT',
C$_6$H$_{12}$N$_4$]) can be formed through the impact of energetic
photons or cosmic rays on the ices, as many laboratory studies have
shown (e.g.  \citealt{sch93}, \citealt{ber95}, \citealt{gre95},
\citealt{ger96}, \citealt{moo98}, \citealt{pal00}, \citealt{mun04}).
Thus far, only the simple species H$_2$O, CO, CO$_2$, CH$_4$,
CH$_3$OH, NH$_3$ and the $^{13}$CO and $^{13}{\rm CO}_2$ isotopes have
been positively identified in the ices toward both low and high mass
protostars as well as extincted background stars whose lines of sight
trace quiescent dense cloud material.  Reasonable evidence exists for
the presence of HCOOH, OCS and the ions NH$_4^+$, OCN$^-$, and
HCOO$^-$ as well, although their existence is sometimes debated
because of inaccurate fits with laboratory spectra or the lack of
multiple bands for independent confirmation.  The identification of
the ions in the ices has been particularly controversial. The first
ion claimed was OCN$^-$ \citep{gri87} and was thought to be produced
by heavy energetic processing of CO:NH$_3$ ices. Later it was realized
that acid-base chemistry in a HNCO:NH$_3$ ice could yield the same
products \citep{nov01}.  Acid-base reactions are very efficient and
occur at temperatures as low as 10 K \citep{rau04b}. The created ions
are less volatile than neutral species and would be able to form an
interstellar salt after the other species have sublimated.  The
presence of complex species formed by ultraviolet (UV) photon and
cosmic ray processing of the simple ices is at least as controversial.
Many claims were made \citep{gre95,gib02}, but the observational
evidence is not firm.  It is crucial to determine the complexity of
the ices in the circumstellar environment of low mass Young Stellar
Objects (YSOs), as they may be delivered directly to comets and they
may ultimately serve as the source of volatiles in planets.  At
sublimation fronts in the warm inner regions of disks they are the
starting point of a complex gas phase chemistry.  The relative
abundances of N-, C-, and O-bearing species coming from the ices
determine directly the type of chemistry in hot cores (or `hot
corinos' for low mass objects; e.g.  \citealt{caz03}).

As a step toward better characterizing the molecular content of icy
grain mantles, this work presents spectra of a large sample of low
mass protostars and background stars obtained with the InfraRed
Spectrometer (IRS; \citealt{hou04}) at the Spitzer Space Telescope
\citep{wer04} in the 5-35 \mum\ wavelength range.  These data are
complemented with spectra at 2-4 \mum\ obtained with ground-based
facilities. Such full coverage over the near and mid-infrared
wavelength ranges is essential in determining the composition of the
ices; most species have several vibrational modes and more secure
identifications can be made when the features are studied
simultaneously. The identification process is aided by studying large
samples of sight-lines facilitating correlation of band strengths and
shapes with each other and, if known, with physical characteristics
along the sight-lines, such as luminosity and evolutionary stage of
any heating source.  The great sensitivity of the IRS allows for
observations of objects with luminosities that are down by orders of
magnitude with respect to what could be observed before in the
mid-infrared (typically 1 L$_{\odot}$ versus $10^4$ L$_{\odot}$).

Most data presented in this paper were obtained from spectral surveys
of nearby molecular clouds and isolated dense cores in the context of
the Spitzer Legacy Program `From Molecular Cores to Planet-Forming
Disks' (`c2d'; \citealt{eva03}). Initial results from this program
emphasize the importance of thermal processing in the evolution of the
ices, as the ices surrounding the low mass YSO HH~46 IRS are more
processed compared to B5 IRS1 \citep{boo04_2}. Ices toward the edge-on
disk CRBR 2422.8-3423 were presented in \citet{pon05} and also show
signs of heating. Finally, mid-infrared spectra of ices toward
background stars tracing quiescent cloud material were published in
\citet{kne05}, showing for the first time that the (unknown) carrier
of the 6.85 \mum\ absorption band is abundant even at the coldest
conditions away from star formation.

This paper specifically addresses questions on the identification of
the 6.0 and 6.85 \mum\ bands. These prominent absorption features are
still not fully identified, despite being detected nearly 30 years ago
with the Kuiper Airborne Observatory \citep{pue79} and commonly
observed toward massive YSOs with the Infrared Space Observatory (ISO;
\citealt{sch96}, \citealt{kea01b}).  The peak position of the 6.85
\mum\ band varies dramatically, and is thought to be a function of the
ice temperature \citep{kea01b}. Either one carrier with a pronounced
temperature dependence of its absorption bands, or two independent
carriers, one much more volatile than the other, must be responsible
for the 6.85 \mum\ feature.  If the former is the case, the ammonium
ion (NH$_4^+$) is considered a promising candidate \citep{sch03}.  The
6.0 \mum\ band was initially thought to be mainly due to the bending
mode of H$_2$O (\citealt{tie84}; however, see \citealt{cox89}), but
analysis of ISO spectra indicated that this is not the case toward
several massive YSOs \citep{sch96, kea01b}. The depth of the 6.0 \mum\
band is in some cases significantly (factor of 2) deeper than that
expected from the 3.0 \mum\ H$_2$O stretching mode.  Part of this
`excess' might be due to a strong vibration of the HCOOH molecule.
Recent observations of background stars show a small excess
($\leq$25\%), of which half could be due to the enhanced band strength
of the H$_2$O bending mode in CO$_2$-rich ices \citep{kne05}.
Finally, it was claimed that much of the 6.0 \mum\ excess and part of
the 6.85 \mum\ band is due to highly processed ices, i.e. a mixture of
complex species with C--H and O--H bonds produced after irradiation of
simple ices \citep{gib02}. In this case, one would expect the 6.0
\mum\ excess to be enhanced in high radiation environments, such as
near more evolved protostars.  Thus, fundamental questions remain on
the identification of the 6.0 and 6.85 \mum\ bands, and with the large
sample presented in this work the possible answers can be further
constrained.  Subsequent papers will specifically address the CO$_2$
\citep{pon08}, CH$_4$ \citep{obe08}, and NH$_3$ (S.  Bottinelli et
al., in preparation) species. Further papers in this series will
investigate the ices toward background stars behind large clouds (C.
Knez et al., in preparation) and isolated cores (A. C. A. Boogert et
al., in preparation).

In \S\ref{sec:sou} the source sample is presented, while the
observations and data reduction of the ground-based L-band and Spitzer
5-38 \mum\ spectra are described in \S\ref{sec:obs}. In
\S\ref{sec:res} the continuum level for the multitude of absorption
features is determined, and subsequently the best value for the H$_2$O
column density derived, followed by an empirical decomposition of the
5-7 \mum\ absorption complex. The positively identified species, or
correlations of the components with other observables are discussed in
\S\ref{sec:id}. Finally, the abundances of the simple species, and
constraints on the carriers of unidentified components are discussed
in \S\ref{sec:dis}.


\begin{deluxetable*}{lllllrlll}
\tabletypesize{\scriptsize}
\tablecolumns{9}
\tablewidth{0pc}
\tablecaption{Source Sample~\label{t:sample}}
\tablehead{
\colhead{Source}& \colhead{RA\tablenotemark{a}} & \colhead{Dec\tablenotemark{a}} & \colhead{Cloud} & \colhead{Type\tablenotemark{b}} & \colhead{$\alpha_{\rm IR}$\tablenotemark{c}} & \colhead{Obs. ID\tablenotemark{d}} & \colhead{Module\tablenotemark{l}} & \colhead{L-band\tablenotemark{m}}\\
\colhead{      }& \colhead{J2000}        & \colhead{J2000}         & \colhead{   }   & \colhead{ } & \colhead{ } & \colhead{ } & \colhead{ } & \colhead{ }\\}
\startdata
L1448 IRS1      & 03:25:09.44            &  +30:46:21.7            &   Perseus       & low         & 0.34                 & \dataset{0005656832} & SL,SH,LH    & NIRSPEC   \\
IRAS 03235+3004 & 03:26:37.45            &  +30:15:27.9            &   Perseus       & low         & 1.44                 & \dataset{0009835520} & SL,LL       & NIRSPEC   \\
IRAS 03245+3002 & 03:27:39.03            &  +30:12:59.3            &   Perseus       & low         & 2.70                 & \dataset{0006368000} & SL,SH       &           \\
L1455 SMM1      & 03:27:43.25            &  +30:12:28.8            &   Perseus       & low         & 2.41                 & \dataset{0015917056} & SL,SH,LL1   &           \\
RNO 15          & 03:27:47.68            &  +30:12:04.3            &   Perseus       & low         &$-$0.21               & \dataset{0005633280} & LL1,SL,SH,LH& NIRSPEC   \\
L1455 IRS3      & 03:28:00.41            &  +30:08:01.2            &   Perseus       & low         & 0.98                 & \dataset{0015917568} & SL,SH,LL1   &           \\
IRAS 03254+3050 & 03:28:34.51            &  +31:00:51.2            &   Perseus       & low         & 0.90                 & \dataset{0011827200} & LL1,SL,SH,LH& NIRSPEC   \\
IRAS 03271+3013 & 03:30:15.16            &  +30:23:48.8            &   Perseus       & low         & 2.06                 & \dataset{0005634304} & LL1,SL,SH,LH& NIRSPEC   \\
IRAS 03301+3111 & 03:33:12.85            &  +31:21:24.2            &   Perseus       & low         & 0.51                 & \dataset{0005634560} & SL,SH,LH    & NIRSPEC   \\
B1-a            & 03:33:16.67            &  +31:07:55.1            &   Perseus       & low         & 1.87                 & \dataset{0015918080} & SL,SH,LL1   & NIRSPEC   \\
B1-c            & 03:33:17.89            &  +31:09:31.0            &   Perseus       & low         & 2.66                 & \dataset{0013460480} & SL,SH,LL1   &           \\
B1-b            & 03:33:20.34            &  +31:07:21.4            &   Perseus       & low         & 0.68                 & \dataset{0015916544} & SL,LL       &           \\
B5 IRS3         & 03:47:05.45            &  +32:43:08.5            &   Perseus       & low         & 0.51\tablenotemark{k}& \dataset{0005635072} & LL1,SL,SH,LH& NIRSPEC   \\
B5 IRS1\tablenotemark{e} 
                & 03:47:41.61            &  +32:51:43.8            &   Perseus       & low         & 0.78\tablenotemark{k}& \dataset{0005635328} & SL,SH,LH    & NIRSPEC   \\
L1489 IRS\tablenotemark{f}       
                & 04:04:43.07            &  +26:18:56.4            &   Taurus        & low         & 1.10                 & \dataset{0003528960} & SL,SH,LH    & NIRSPEC   \\
IRAS 04108+2803B\tablenotemark{f}
                & 04:13:54.72            &  +28:11:32.9            &   Taurus        & low         & 0.90                 & \dataset{0003529472} & SL,SH,LH    & NIRSPEC   \\ 
HH~300\tablenotemark{f}          
                & 04:26:56.30            &  +24:43:35.3            &   Taurus        & low         & 0.79                 & \dataset{0003530752} & SL,SH,LH    & NIRSPEC   \\       
DG Tau B\tablenotemark{f}        
                & 04:27:02.66            &  +26:05:30.5            &   Taurus        & low         & 1.16                 & \dataset{0003540992} & SL,SH,LH    & NIRSPEC   \\
HH~46 IRS\tablenotemark{e} 
                & 08:25:43.78            &$-$51:00:35.6            &   HH~46         & low         & 1.70                 & \dataset{0005638912} & SL,SH,LH    & ISAAC  \\
IRAS 12553-7651 & 12:59:06.63            &$-$77:07:40.0            &   Cha           & low         & 0.76                 & \dataset{0009830912} & LL1,SL,SH,LH&           \\
IRAS 13546-3941 & 13:57:38.94            &$-$39:56:00.2            &   BHR92         & low         & $-$0.06              & \dataset{0005642752} & SL,SH,LH    &           \\
IRAS 15398-3359 & 15:43:02.26            &$-$34:09:06.7            &   B228          & low         & 1.22                 & \dataset{0005828864} & SL,SH,LL1   &           \\
Elias 29\tablenotemark{g}        
                & 16:27:09.42            &$-$24:37:21.1            &   Oph           & low         & 0.53\tablenotemark{k}& 26700814             & SWS01 sp3   &             \\
CRBR 2422.8-3423\tablenotemark{h}
                & 16:27:24.61            &$-$24:41:03.3            &   Oph           & low         & 1.36                 & \dataset{0009346048} & SL,SH,LH    & NIRSPEC   \\
RNO 91          & 16:34:29.32            &$-$15:47:01.4            &   L43           & low         & 0.03                 & \dataset{0005650432} & SL,SH,LH    & ISAAC   \\
IRAS 17081-2721 & 17:11:17.28            &$-$27:25:08.2            &   B59           & low         & 0.55                 & \dataset{0014893824} & SL,SH,LL1   & NIRSPEC   \\
SSTc2dJ171122.2-272602
                & 17:11:22.16            &$-$27:26:02.3            &   B59           & low         & 2.26                 & \dataset{0014894336} & SL,LL       &           \\
2MASSJ17112317-2724315 
                & 17:11:23.13            &$-$27:24:32.6            &   B59           & low         & 2.48                 & \dataset{0014894592} & SL,LL       & NIRSPEC   \\
EC 74           & 18:29:55.72            &  +01:14:31.6            &   Serpens       & low         & $-$0.25              & \dataset{0009407232} & SL,SH,LH    & NIRSPEC   \\
EC 82           & 18:29:56.89            &  +01:14:46.5            &   Serpens       & low         & 0.38\tablenotemark{k}& \dataset{0009407232} & SL,SH,LH    &         \\
SVS 4-5         & 18:29:57.59            &  +01:13:00.6            &   Serpens       & low         & 1.26                 & \dataset{0009407232} & SL,SH,LH    & ISAAC   \\
EC 90           & 18:29:57.75            &  +01:14:05.9            &   Serpens       & low         & $-$0.09              & \dataset{0009828352} & SL,SH,LH    &           \\    
EC 92           & 18:29:57.88            &  +01:12:51.6            &   Serpens       & low         & 0.91                 & \dataset{0009407232} & SL,SH,LH    & NIRSPEC   \\
CK4             & 18:29:58.21            &  +01:15:21.7            &   Serpens       & low         &$-$0.25               & \dataset{0009407232} & SL,SH,LH    &           \\
R CrA IRS 5     & 19:01:48.03            &$-$36:57:21.6            &   CrA           & low         & 0.98                 & \dataset{0009835264} & SL,SH,LL1   & ISAAC     \\
HH 100 IRS\tablenotemark{i}      
                & 19:01:50.56            &$-$36:58:08.9            &   CrA           & low         & 0.80                 & 52301106             & SWS01 sp4   &             \\
CrA IRS7 A      & 19:01:55.32            &$-$36:57:22.0            &   CrA           & low         & 2.23                 & \dataset{0009835008} & SL,SH,LH    & ISAAC     \\
CrA IRS7 B      & 19:01:56.41            &$-$36:57:28.0            &   CrA           & low         & 1.63                 & \dataset{0009835008} & SL,SH,LH    & ISAAC     \\
CrA IRAS32      & 19:02:58.69            &$-$37:07:34.5            &   CrA           & low         & 2.15                 & \dataset{0009832192} & SL,SH,LL1   &           \\
L1014 IRS       & 21:24:07.51            &  +49:59:09.0            &   L1014         & low         & 1.28                 & \dataset{0012116736} & SL,LL       & NIRSPEC   \\
IRAS 23238+7401 & 23:25:46.65            &  +74:17:37.2            &   CB 244        & low         & 0.95                 & \dataset{0009833728} & SL,SH,LH    & NIRSPEC   \\
                &                        &                         &                 &             &                      &                      &             &           \\
W3 IRS5\tablenotemark{i}         
                & 02:25:40.54            &$+$62:05:51.4            &                 & high        & 3.53                 & 42701302             & SWS01 sp3   &             \\ 
MonR2 IRS3\tablenotemark{i}      
                & 06:07:47.8             &$-$06:22:55.0            &                 & high        & 1.66                 & 71101712             & SWS01 sp3   &             \\
GL989\tablenotemark{i}           
                & 06:41:10.06            &$+$09:29:35.8            &                 & high        & 0.52                 & 71602619             & SWS01 sp3   &             \\
W33A\tablenotemark{i}            
                & 18:14:39.44            &$-$17:52:01.3            &                 & high        & 1.92                 & 32900920             & SWS01 sp4   &             \\ 
GL7009S\tablenotemark{i}         
                & 18:34:20.91            &$-$05:59:42.2            &                 & high        & 2.52                 & 15201140             & SWS01 sp3   &             \\
GL2136\tablenotemark{i}          
                & 18:22:26.32            &$-$13:30:08.2            &                 & high        & 1.48                 & 33000222             & SWS01 sp3   &             \\
S140 IRS1\tablenotemark{i}       
                & 22:19:18.17            &$+$63:18:47.6            &                 & high        & 1.57                 & 22002135             & SWS01 sp4   &             \\
NGC7538 IRS9\tablenotemark{i}    
                & 23:14:01.63            &$+$61:27:20.2            &                 & high        & 2.31                 & 09801532             & SWS01 sp2   &             \\
                &                        &                         &                 &             &                      &                      &             &           \\
Elias 16\tablenotemark{k}        
                & 04:39:38.88  	   &$+$26:11:26.6            &   Taurus        & bg          & -                    & \dataset{0005637632} & SL,SH       &          \\
EC 118\tablenotemark{k}          
                & 18:30:00.62  	   &$+$01:15:20.1            &   Serpens       & bg          & -                    & \dataset{0011828224} & SL,SH       &          \\
\enddata
\tablenotetext{a}{Position used in Spitzer/IRS observations}
\tablenotetext{b}{Source type: 'low'=low mass YSO, 'high'=massive YSO, 'bg'=background star}
\tablenotetext{c}{Broad-band spectral index as defined in
Eq.~\ref{eq:alpha}}
\tablenotetext{d}{AOR key for Spitzer and TDT number for ISO observations}
\tablenotetext{e}{Published previously in Boogert et al. 2004}
\tablenotetext{f}{Published previously in Watson et al. 2004}
\tablenotetext{g}{Published previously in Boogert et al. 2000}
\tablenotetext{h}{Published previously in Pontoppidan et al. 2005}
\tablenotetext{i}{Published previously in Keane et al. 2001}
\tablenotetext{j}{Published previously in Knez et al. 2005}
\tablenotetext{k}{$\alpha$ enhanced due to  foreground extinction. Exclusion $K_{\rm s}$-band flux gives much lower $\alpha$: $-$0.16 (Elias 29), $-0.02$ (B5 IRS1), 0.18 (B5 IRS3), and 0.38 (EC 82)}
\tablenotetext{l}{Spitzer/IRS modules used: SL=Short-Low (5-14 \mum, $R\sim100$), LL=Long-Low (14-34 \mum, $R\sim100$), SH=Short-High (10-20 \mum, $R\sim600$), LH=Long-High (20-34 \mum, $R\sim600$); ISO SWS modes used (2.3-40 \mum): SWS01 speed 1 ($R\sim250$), speed 2 ($R\sim250$), speed 3 ($R\sim400$), speed 4 ($R\sim800$)}
\tablenotetext{m}{Complementary ground-based L-band observations with Keck/NIRSPEC or VLT/ISAAC}
\end{deluxetable*}

\section{Source Sample}~\label{sec:sou}

The source sample is selected based on the presence of ice absorption
features and consists of a combination of known low mass YSOs and new
ones identified from their Spitzer/IRAC and MIPS broad-band SEDs
(Table~\ref{t:sample}; \citealt{eva07}).  For the known objects, their
mid-infrared (5-20 \mum) spectra are presented here for the first
time, with the exception of B5 IRS1, HH~46 IRS \citep{boo04_2, nor04},
CRBR 2422.8-3423 \citep{pon05}, and L1489 IRS, IRAS 04108+2803B, HH
300, and DG Tau B \citep{wat04}.  The YSOs are located in the Perseus,
Taurus, Serpens, and Corona Australis molecular cloud complexes, and a
number of nearby isolated dense cores (Table~\ref{t:sample}). YSOs in
the Ophiuchus cloud are notably absent, because the Spitzer/IRS
spectra were not available at the time this paper was written.  The
spectral energy distributions of the sample of 41 low-mass YSOs span a
wide range of spectral indices $\alpha=-0.25$ to +2.70, with $\alpha$
defined as

\begin{equation}
\alpha=\frac{d log(\lambda F_\lambda)}{d log(\lambda)} ~\label{eq:alpha}
\end{equation}

\noindent including all photometry ($F_\lambda$) available between the
$\lambda=$ 2.17 \mum\ $K_{\rm s}$-band from the Two Micron All Sky
Survey (2MASS; \citealt{skr06}) and the $\lambda=$ 24 \mum\
Spitzer/MIPS band.  In the infrared broad-band classification scheme
\citep{wil01}, most objects (35 out of 41) fall in the embedded Class
0/I category ($\alpha>$0.3).  The remaining 6 objects are Flat-type
objects ($-0.3<\alpha<0.3$; \citealt{gre94}).  Although most objects
fall in the Class 0/I category, the dispersion within it is
large. Some objects are envelope-dominated Class 0 sources undetected
in the 2MASS $K_{\rm s}$-band (e.g. L1455 SMM1, B1-c), while others
are more likely to be Flat-type sources with dispersed envelopes
and/or face-on disks extincted by foreground dust.  The latter sources
were identified by excluding the $K_{\rm s}$-band flux in the $\alpha$
determination, and are marked as such in Table~\ref{t:sample}. Thus,
our sample represents a wide range of YSO evolutionary stages, and
probably source orientation.

To contrast the possibly processed ices toward YSOs with ices toward
quiescent regions, published mid-infrared spectra of 2 background
stars toward the Serpens and Taurus clouds were included in the
analysis \citep{kne05}.  In addition, 8 massive YSOs and 2 low mass
YSOs observed with the ISO satellite were included in the sample
(Table~\ref{t:sample}). These serve as a reference in the ice feature
analysis, because their mid-infrared spectra are well-studied at
higher spectral resolution than is possible with Spitzer/IRS and
because the massive YSOs may trace different physical conditions that
may affect the formation and evolution of the ices.


\section{Observations and Data Reduction}~\label{sec:obs}

Spitzer/IRS spectra were obtained as part of the c2d Legacy program
(PIDs 172 and 179) as well as a dedicated open time program (PID
20604). In addition, several previously published GTO spectra were
included \citep{wat04}.  The coordinates and IRS modules that were
used for each source are listed in Table~\ref{t:sample}.  For all
modules the 2-dimensional Basic Calibrated Data (BCD) spectral images
produced by the SSC pipeline version S13.2.0 or later were used as the
starting point of the reduction. The low resolution `Short-Low' and
`Long-Low' modules (SL and LL; $R=\lambda/\Delta\lambda=60-120$) were
reduced in a way that is customary for ground-based spectra. First,
the spectral and spatial dimensions were orthogonalized, and then the
2-dimensional images of the two nodding positions were subtracted in
order to remove extended emission. In some cases the background
emission is highly structured or multiple sources are present in the
slit, and instead of subtracting nodding pairs, the background
emission was determined from neighboring columns in the extraction
process.  A fixed width extraction was performed and the 1-dimensional
spectra were then averaged. Subsequently the spectra were divided by
spectra of the standard stars HR 2194 (A0~V; PID 1417; AOR keys
\dataset{0013024512} and \dataset{0013024768} for SL1 and SL2
respectively) and HR 6606 (G9~III; PID 1421; AOR key
\dataset{0013731840} for LL) reduced in the same way in order to
correct for wavelength-dependent slit losses. Spectral features were
divided out using the photospheric models of \citet{dec04}.  The
spectra of the `Short-High' and `Long-High' modules (SH and LH;
$R\sim$600) were reduced using the c2d pipeline \citep{lah07}. An
important difference with the low resolution modules is the need for
the removal of spectral fringes, which is done in the c2d pipeline as
well. Many of the Spitzer spectra were complemented by ground-based
Keck/NIRSPEC \citep{mcl98} and VLT/ISAAC \citep{moo97} L-band spectra
at resolving powers of $R$=2000 and 600 respectively. These were
reduced in a way standard for ground-based long-slit spectra.
Finally, the Spitzer/IRS modules and ground-based spectra were
multiplied along the flux scale in order to match Spitzer/IRAC 3.6,
4.5, 5.8, and 8.0 \mum\ and Spitzer/MIPS 24 \mum\ photometry
\citep{eva07}, using the appropriate filter profiles.

For the ISO/SWS satellite spectra used in this paper, the latest SPD
pipeline version 10.1 products were taken from the ISO archive and the
detector scans were cleaned from cosmic ray hits and averaged. The
final spectra do not show any significant differences with respect to
the data published in \citet{kea01b}.

\section{Results}~\label{sec:res}

\begin{figure*}
\includegraphics[angle=90, scale=1.33]{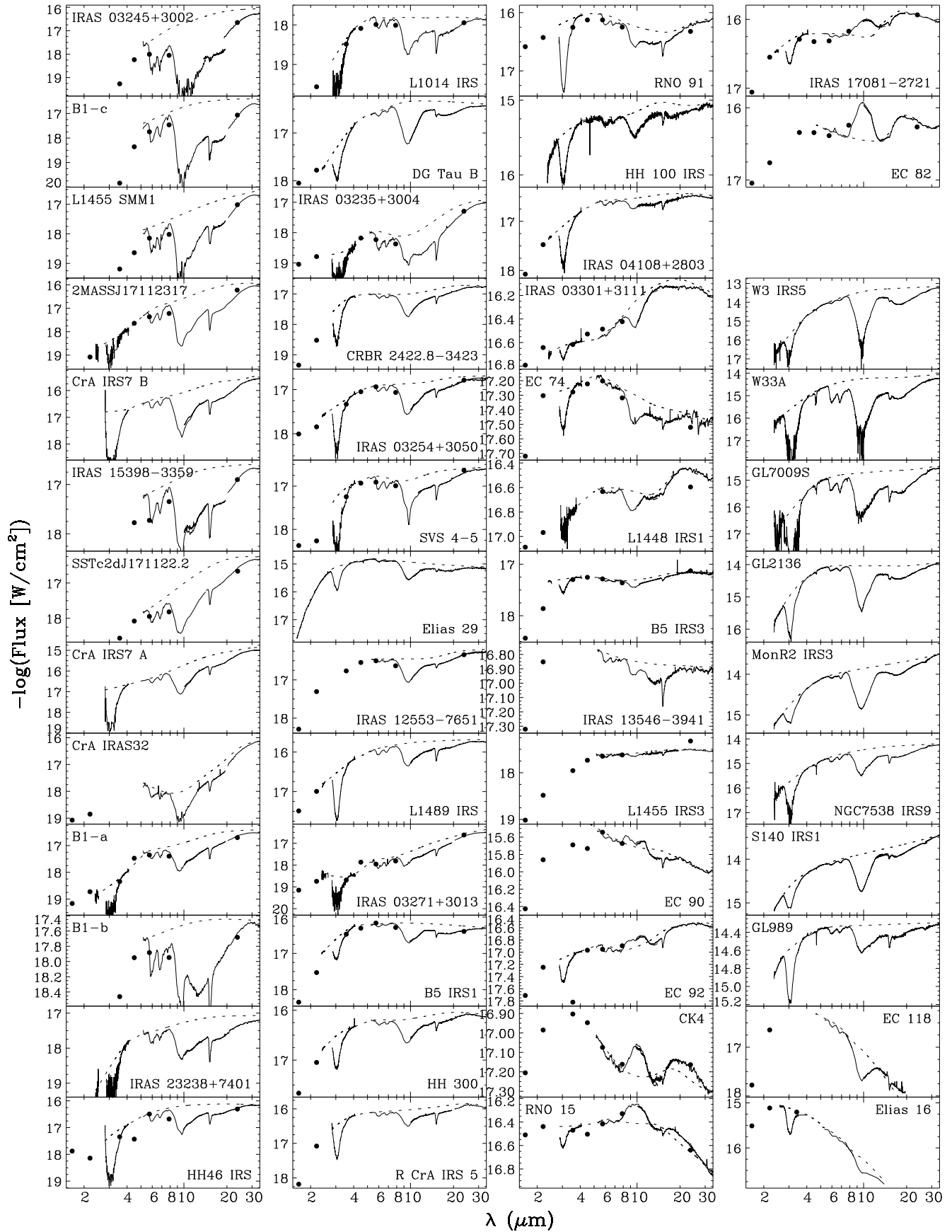}
\caption{Observed Spitzer/IRS and complementary ISO/SWS and ground
   based L-band spectra and the adopted continua (dotted lines).  The
   large dots represent 2MASS and Spitzer/IRAC and MIPS photometry
   from the c2d catalogs \citep{eva07}. Sources are ordered from top
   to bottom and left to right in decreasing 9.7 \mum\ absorption band
   depth. Massive YSOs and background stars are separated in the right
   panel below the white space.}~\label{f:sampl}
\end{figure*}

All sources in the sample show numerous absorption features on top of
rising, flat, or falling 2--35 \mum\ Spectral Energy Distributions
(Fig.~\ref{f:sampl}).  Most of these features are due to ices
(\S\ref{sec:abs}), and they are often accompanied by strong 9.7 and
weaker 18 \mum\ silicate absorption bands; but toward some sources,
the silicate bands are in emission (e.g. RNO 15, IRAS 17081-2721).
Also, prominent narrow emission lines of H$_2$ (S[5], S[6], and S[7])
are observed toward IRAS 03271+3013, CrA IRAS32, and L1455 SMM1.
These belong to the most deeply embedded objects ($\alpha>2$;
Table~\ref{t:sample}) in the sample and the line emission is most
likely related to molecular outflow activity (e.g.  \citealt{nor04};
\citealt{neu06}).  CrA IRAS32 shows the shock-related 5.340 \mum\
emission line of Fe$^+$ \citep{neu06}, and a strong line at 6.628
\mum, perhaps also due to Fe$^+$.  These emission lines are not
further analyzed here, but their presence compromises the analysis of
the underlying absorption features in these particular sources. H~{\sc
i} emission lines (most notably Br $\alpha$ at 4.052 \mum\ and Pf
$\gamma$ at 3.740 \mum) are present in many of the ground-based L-band
spectra, but at the high spectral resolution contamination with the
ice absorption features is not a problem. In sources with strong
L-band H~{\sc i} emission (e.g. RNO 91) the blended Pf $\alpha$ (7.459
\mum) and Hu $\beta$ (7.502 \mum) lines are weakly visible in the IRS
spectra ($\sim$3\% of the continuum), affecting the continuum
determination for the low contrast ice features in that region (\S
\ref{sec:abs}; \citealt{obe08}).

\begin{figure*}
\includegraphics[angle=90, scale=1.35]{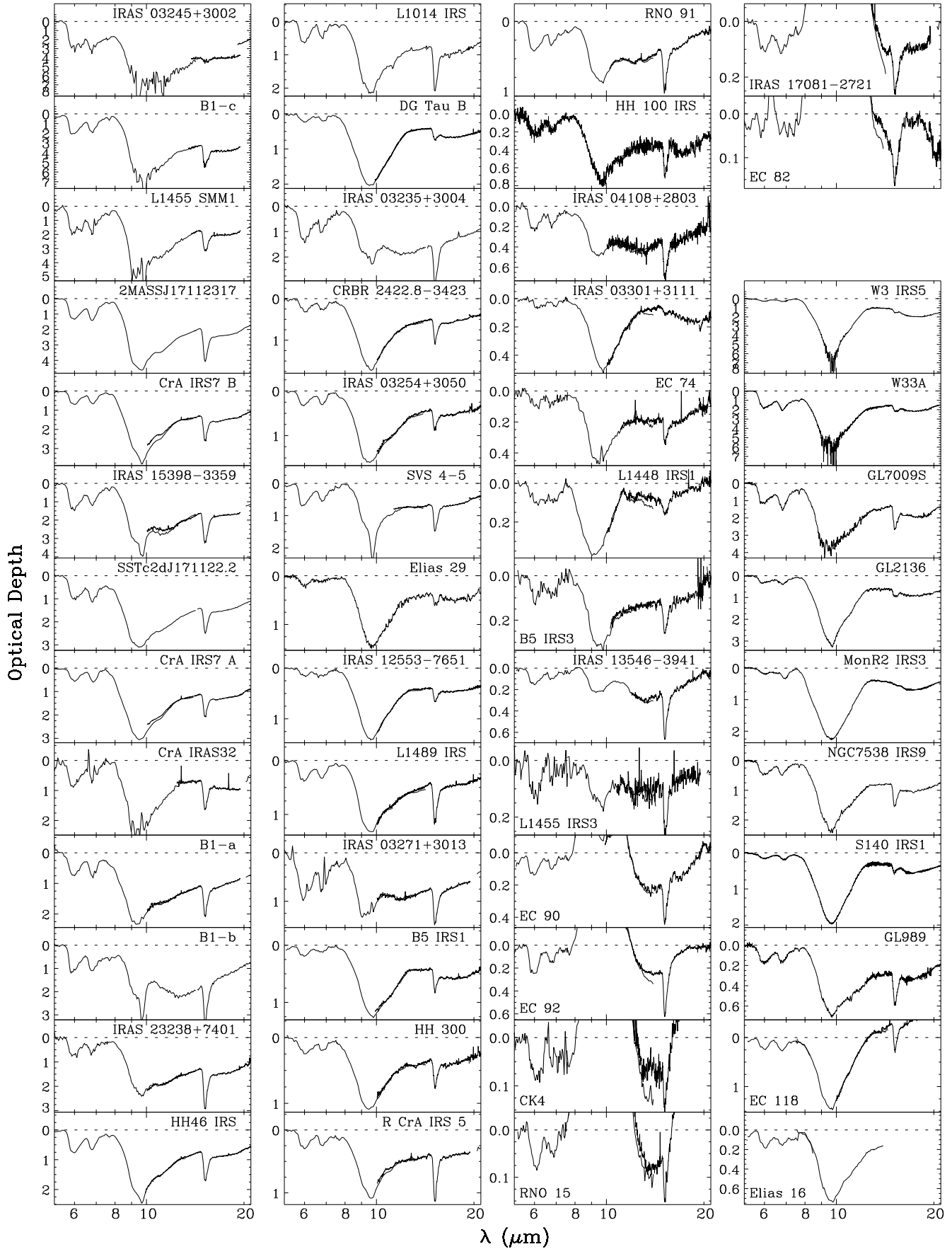}
\caption{Optical depth 5-20 \mum\ spectra, ordered from top to bottom
  and left to right in decreasing 9.7 \mum\ absorption band depth.
  The horizontal dotted line indicates zero optical depth for
  reference.  Massive YSOs and background stars are separated in the
  right panel below the white space.}~\label{f:tau}
\end{figure*}

\begin{figure*}
\center
\includegraphics[angle=90, scale=0.55]{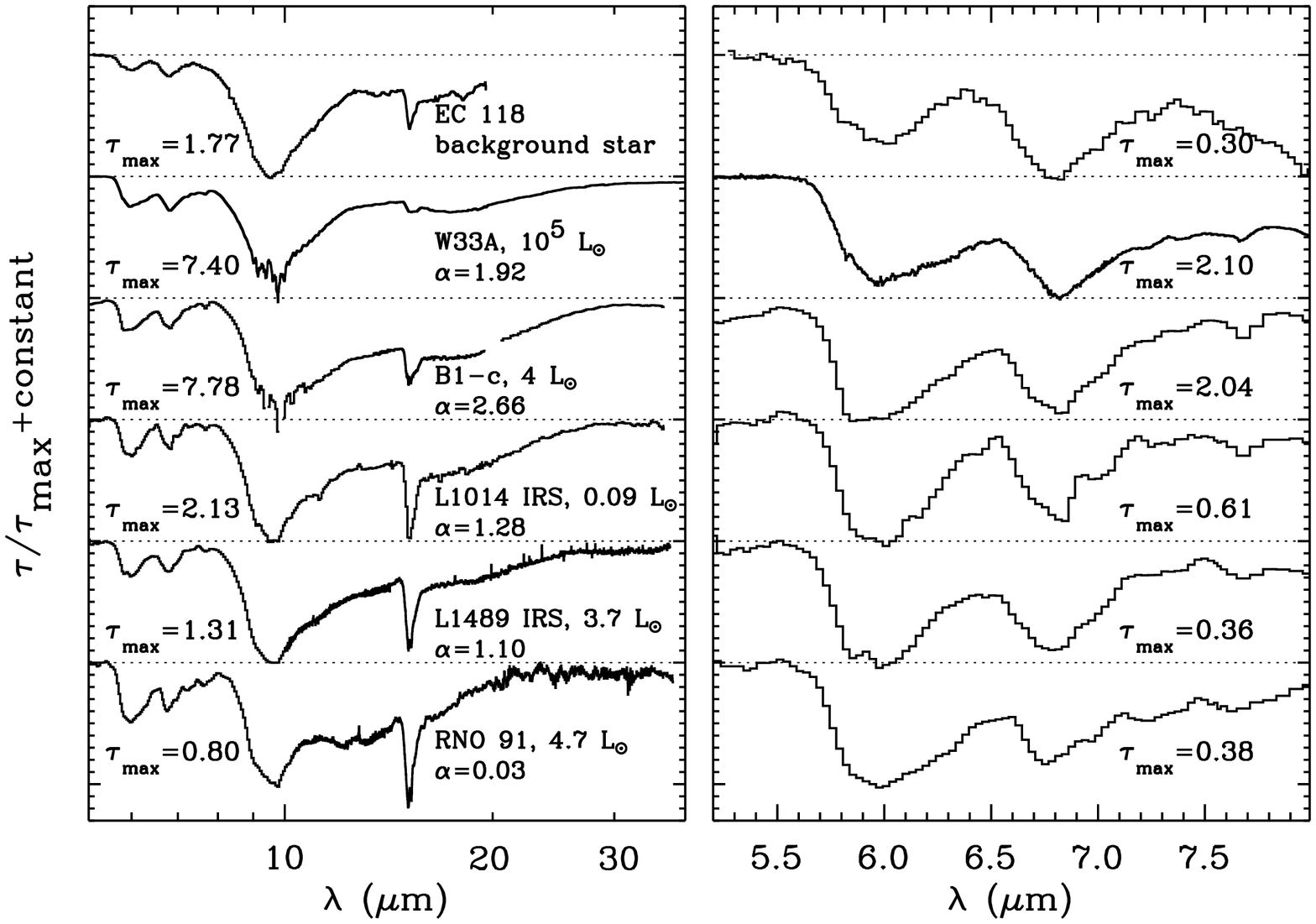}
\caption{Optical depth spectra between wavelengths of 5.2-38 \mum\
  (left panel), and 5.2-8.0 \mum\ (right panel) of a wide variety of
  sight-lines. From top to bottom: EC 118, a background star behind the
  Serpens cloud \citep{kne05}, W 33A, a massive YSO \citep{kea01b}, and
  the low mass YSOs B1-c in the Perseus cloud, L1014 IRS in the
  isolated core L1014, L1489 IRS in the Taurus cloud, and RNO 91 in
  the L43 core near the Ophiuchus cloud. The low mass YSOs are ordered
  from top to bottom in decreasing 2-25 \mum\ SED spectral index
  $\alpha$ (Eq.~\ref{eq:alpha}) as indicated in the left panel. Also
  indicated is the bolometric luminosity. The spectra are scaled to
  the peak optical depth $\tau_{\rm max}$ in the plotted range
  indicated near each spectrum. Note that despite the very different
  nature of the sight-lines, the same ice absorption features are
  observed, although the shapes and depths vary from source to
  source.}~\label{f:opdepth}
\end{figure*}

\subsection{Continuum Determination}~\label{sec:cont}

Determining the baseline for the absorption features in the spectra of
embedded protostars is not trivial.  The continuum SED underlying the
many blended absorption features originates from warm dust near the
central star and is dependent on poorly characterized parameters such
as disk and envelope mass, size, and inclination, and dust composition
\citep{whi03}.  The continuum applied to each source and indicated in
Fig.~\ref{f:sampl} was derived as follows.  Short-ward of 5 \mum\ the
continuum is much better defined than at longer wavelengths, and
therefore the 2-5 and 5-35 \mum\ regions are treated separately.  For
the short wavelength part, continuum points selected at 2.3, 3.9, and
5.0 \mum\ are interpolated with a smooth spline function.  Inclusion
of the 2.3 \mum\ portion of the spectrum, or $K_{\rm s}$-band
photometry, is essential in choosing the best continuum for the 3.0
\mum\ absorption band. The continuum long-ward of 5 \mum\ is much less
constrained, because the entire 5-32 \mum\ region is covered by
absorption features.  Even the `emission peak' at 8.0 \mum\ is covered
by the long wavelength wing of the 6.0 \mum\ H$_2$O ice absorption
band (e.g.  \citealt{hud93}), and the short wavelength wing of the 9.7
\mum\ silicate band.  The latter often extends toward shorter
wavelengths compared to the silicate spectrum observed in the diffuse
medium (e.g.  toward GCS 3; \citealt{kem04}).  Some sources also show
an additional depression at 8 \mum\ due to yet unknown species; the
best known example of this is the massive YSO W33A \citep{kea01b,
  gib02}.  Thus, an interpolation over a large wavelength range is
required, with few constraints.  When possible, a low order polynomial
(on a $log(\lambda F_{\lambda})$ versus $log(\lambda)$ scale) was used
to interpolate between the 5 and 32 \mum\ continuum points.  This
works best for the most deeply embedded sources (e.g.  B1-c,
2MASSJ17112317). For many sources, some guidance for the polynomial
fit is provided by scaling the silicate spectrum of the diffuse medium
source GCS 3 \citep{kem04} to the peak optical depth at 9.7 \mum\ and
by scaling a laboratory H$_2$O ice spectrum to the 3.0 \mum\ band.
Less embedded sources, i.e. the sources with relatively strong
near-infrared emission (e.g.  RNO 91, IRAS 03235+3004), show a
decreasing continuum between 5-10 \mum\ and an emission bump at longer
wavelengths. Here, a best effort spline continuum was defined, guided
by modified black-body curves below 10 \mum. For the background stars
the optical depth spectra presented in \citet{kne05} were used.

\subsection{Absorption Features}~\label{sec:abs}

The derived continua were used to put the spectra on an optical depth
scale for the analysis of the wealth of absorption features
(Figs.~\ref{f:sampl} and~\ref{f:tau}).  The well known 6.0 and 6.85
\mum\ bands (e.g.  \citealt{kea01b}) are detected in all sources,
along with the 9.7 \mum\ band of silicates (either in absorption or
emission) and the 3.0 \mum\ H$_2$O and 15 \mum\ CO$_2$ ice bands when
available. A number of weaker ice features (7.25, 7.40, 7.67, 9.0, and
9.7 \mum) are detected in several sources as well.  Examples in
Fig.~\ref{f:opdepth} show that these same absorption features are
present in sources of a very different nature (background stars,
massive YSOs, Class 0, I, and flat spectrum low mass YSOs), but that
the depths and the profiles vary.  Very deep 6.0 and 6.8 \mum\ bands,
with peak optical depths of up to 2, are observed in the Class 0
sources. These are the deepest ice bands ever detected in low mass
YSOs, comparable to the most obscured massive YSOs (e.g.  W33A, GL
7009S).

\subsubsection{The H$_2$O Ice Column Density}~\label{sec:h2o}

\begin{figure}
\includegraphics[angle=90, scale=0.60]{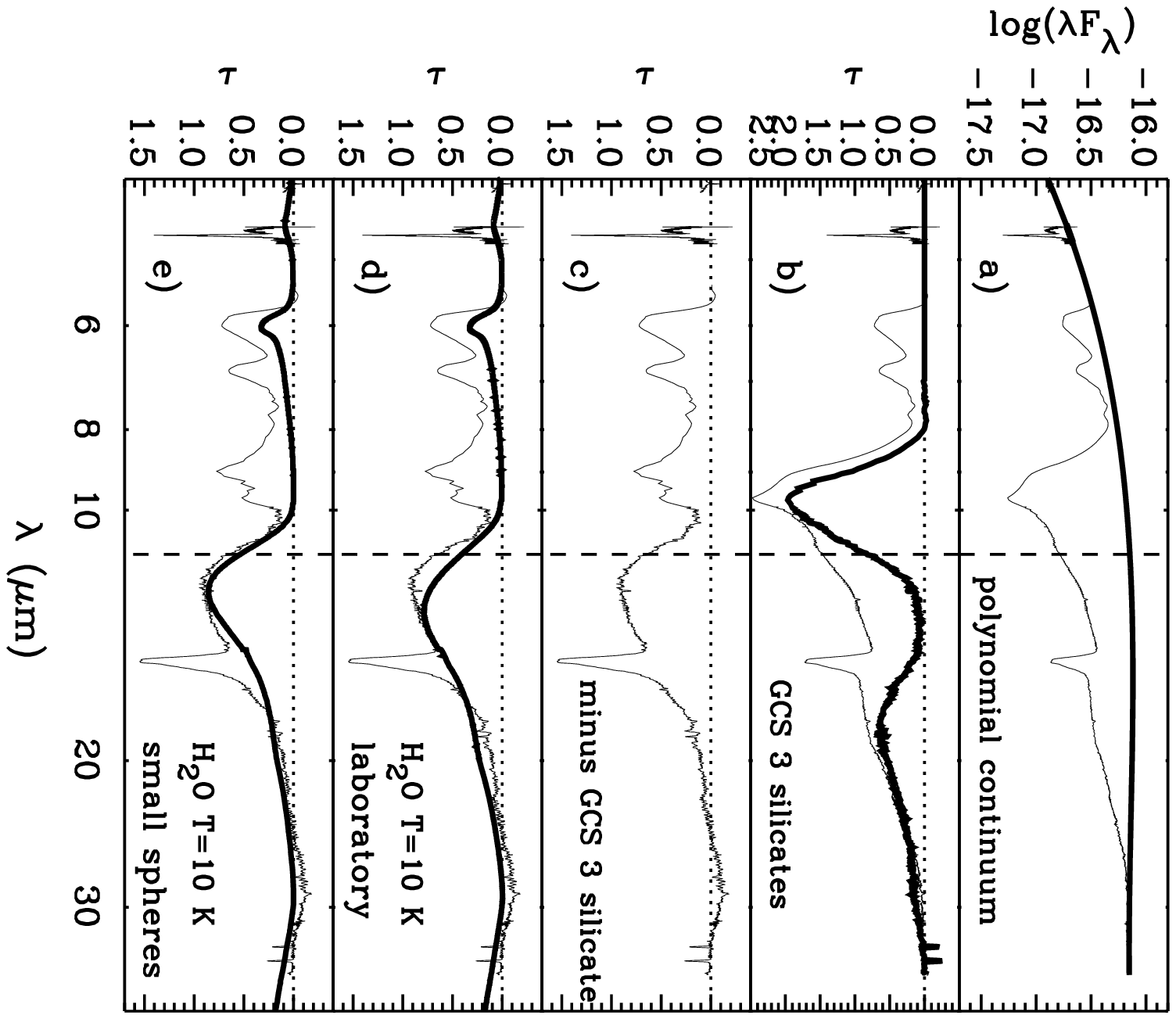}
\caption{Proof of the presence of the 13 \mum\ solid H$_2$O libration
  mode in the source HH~46 IRS. {\bf Panel a} shows a smooth, unbiased
  polynomial continuum (thick line). The optical depth spectrum is
  compared to the silicate spectrum of the galactic center source GCS3
  in {\bf Panel b} (thick line; \citealt{kem04}). After subtracting
  GCS3 a prominent absorption feature in found at 10-20 \mum\ which is
  likely due to the strong libration mode of H$_2$O ({\bf panel c}),
  as is shown by a comparison to a laboratory spectrum of pure
  amorphous H$_2$O ice at $T=$10 K (thick line; {\bf panel d}).  This
  band is particularly sensitive to grain shape effects.  and a better
  fit is obtained with small spherical grains (thick line; {\bf panel
    e}).  The column densities of H$_2$O derived from the 3 and 13
  \mum\ bands are in good agreement, further validating the
  correctness of the applied method. The vertical dashed line is the
  11.3 \mum\ absorption feature identified in \citet{boo04} and
  \citet{kes05}.}~\label{f:lib}
\end{figure}

H$_2$O is the most abundant molecule in interstellar ices (e.g.
\citealt{kea01b,gib04}), and strong, broad absorption bands are
present over much of the observed spectral range: the O-H stretching
mode between $\sim$2.7-3.6 \mum, the O-H bending mode between
$\sim$5.4-9 \mum, and the libration mode between $\sim$10-30 \mum.
Many absorption features of other species overlap with these H$_2$O
bands, especially in the 5-8 \mum\ region (\S\ref{sec:60}), and in
order to separate them, accurate (10-20\%) solid H$_2$O column
densities are required.  The 3.0 \mum\ band provides, in principle,
the most accurate H$_2$O column density, because the O-H stretching
mode is very strong and dominates over other species \citep{tie84}.
Evidently, the ice band must not be saturated, the continuum
signal-to-noise values must be sufficient, and photometry must be
available below the 2.8 \mum\ atmospheric gap for an accurate
continuum determination (\S\ref{sec:cont}). None of these requirements
are met for the most embedded objects in our sample (e.g. B1-c,
2MASSJ17112317-2724315).  For sources that do meet the requirements,
the long wavelength wing that often accompanies the 3.0 \mum\ band is
not taken into account in the H$_2$O column determination.  The width
of a laboratory pure H$_2$O absorption spectrum at the best fitting
temperature is assumed \citep{hud93}, using the integrated band
strength $A=2.0\times 10^{-16}$ cm/molecule \citep{hag81}. Thus, the
calculated column density represents that of absorption by pure H$_2$O
on small grains.  If the wing were taken into account, the H$_2$O
column density would be larger by at most 15\% in some sources.  The
wing, which has an integrated optical depth of at most 30\% with
respect to the rest of the 3 \mum\ band, is probably caused by a
combination of scattering on large ice coated grains (radius $>$0.5
\mum) and absorption by ammonia hydrates \citep{hag83, dar01}.  Only
the former would contribute to the H$_2$O column density, but less
than expected for small grains because the integrated extinction
(absorption+scattering) strength per molecule is a factor of 2 larger
for large grains than it is for small grains (e.g. Fig. 7 in
\citealt{dar01}).

\begin{figure}
\includegraphics[angle=90, scale=0.47]{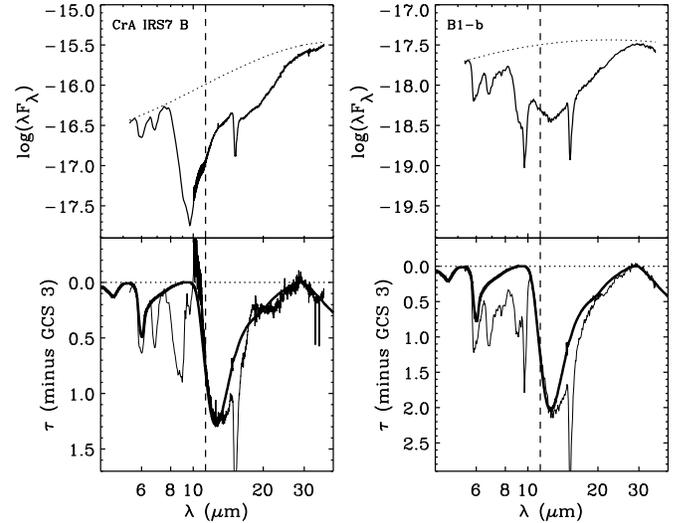}
\caption{In deeply embedded sources the 13 \mum\ H$_2$O libration mode
can be used to reliably measure the solid H$_2$O column density.
'Silicate free' optical depth spectra were derived in the same way as
shown in Fig. \ref{f:lib} for the YSOs CrA~IRS7~B (left panels) and
B1-b (right panels). For each source the top panel shows the observed
spectrum and the polynomial continuum (dotted line), and the bottom
panel the optical depth spectrum.  Note the very prominent libration
mode and the peculiar absence of silicates toward B1-b.  Again, small
spherical grains of pure H$_2$O at $T$=10 K (thick line) fit the peak
position of the observed band well, although excess absorption is
visible on the long wavelength side.  Finally, note that the down-turn
above 30 \mum\ is likely real and due to the onset of the H$_2$O
lattice mode, as the good fits with the laboratory data
show.}~\label{f:libonly}
\end{figure}

An alternative, and sometimes the only, way to determine the H$_2$O
column density is by measuring the peak strength of the libration mode
in the $11-14$ \mum\ region. This band is difficult to isolate in
the spectrum because it is several \mum\ wide and wedged between the
strong 9.7 and 18 \mum\ silicate features.  However, application of a
low order polynomial continuum to HH~46 IRS (Fig.~\ref{f:lib}) and
subtraction of a standard interstellar silicate spectrum (GCS 3;
\citealt{kem04}) shows a broad feature centered on $\sim$12.3 \mum\
that is attributed to the H$_2$O libration mode. Using a laboratory
spectrum of a thin film of pure H$_2$O ice at $T=$10 K \citep{hud93}
the depth is in good agreement with the 3.0 \mum\ stretching mode, but
there is a discrepancy in the peak position by about 1 \mum\
(Fig.~\ref{f:lib}).  The peak position of the libration mode is quite
sensitive to the shape of the grains, and for small spherical grains
(using the formulation of \citealt{boh83}) a much better calculated
fit is obtained (Fig.~\ref{f:lib}). At such large shift to shorter
wavelengths, the H$_2$O libration mode overlaps with the steep edge of
the 9.7 \mum\ silicate band and a substructure is introduced
reminiscent of the 11.3 \mum\ absorption feature reported in
\citet{boo04_2} and \citet{kes05}.  Finally, this method of
determining the H$_2$O column density works well for deeply embedded
sources (Fig.~\ref{f:libonly}), but it fails when a silicate emission
component is present, filling in the H$_2$O absorption (e.g.  RNO 91).

\subsubsection{The 5--8 \mum\ Absorption Complex}~\label{sec:60}

Prominent absorption bands are present at 6.0 and 6.8 \mum\ in all
sources. In-between these bands ($\sim$6.5 \mum), the optical depth is
non-zero and varies relative to the 6.0 and 6.8 \mum\ peak optical
depths for different sources (e.g.  Fig.~\ref{f:opdepth}).  Beyond 7.0
\mum, `featureless' absorption is present up to 8 \mum\ that
approaches zero optical depth at 8 \mum\ in some sources, but
continues to be significant in others and merges with the 9.7 \mum\
silicate band. Here an attempt is made to decompose this complex blend
of absorption features.  Distinct, much weaker bands that are present
in a subset of sources at 7.25, 7.4, and 7.6 \mum\ are discussed
separately (\S\ref{sec:7_8}).

\begin{figure}
\includegraphics[angle=90, scale=0.52]{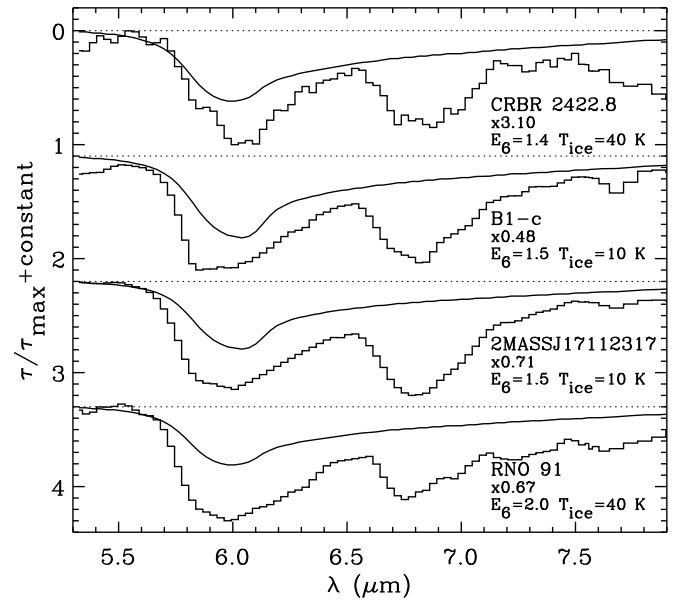}
\caption{A selection of 5--8 \mum\ spectra of four low mass YSOs on
  normalized optical depth scales (histogram line). The normalization
  factors are indicated for each source. The smooth solid line
  represents a laboratory spectrum of pure, amorphous solid H$_2$O at
  the indicated temperature at the column density derived from the 3.0
  and 13 \mum\ bands. Clearly, in all sources a significant fraction
  of the absorption is not explained by solid H$_2$O, and it varies
  from source-to-source. For the 6.0 \mum\ feature this fraction is
  indicated by the value of $E_6$, defined in Eq.~\ref{eq:e6}. Note
  the variations of the profiles of the 6.0 and 6.85 \mum\ bands, as
  well as the presence of weak features at 7.25, 7.4, and 7.65 \mum.}
  ~\label{f:58}
\end{figure}

Solid H$_2$O is a major contributor to the optical depth in the 5--8
\mum\ region by its O-H bending mode at 6.0 \mum\ and the overtone of
the libration mode extending between 6 and 8 \mum\ (e.g.
\citealt{dev01}). For consistency with previous work, although
formally incorrect, these overlapping bands will be referred to as a
single mode, the O-H bending mode of H$_2$O.  Its absorption profile
depends on the ice structure. In particular, as an amorphous ice is
heated from $T=10$ to 80 K, the distinct peak at 6.0 \mum\ becomes
weaker \citep{hud93, mal98}. At higher temperatures, the shape remains
relatively constant. For almost half of the sample the ice temperature
could be determined from the shape of the 3.0 \mum\ H$_2$O stretching
mode (Table~\ref{t:tau}). Most other YSOs are deeply embedded, and
$T=10$ K was assumed.  Subsequently a laboratory spectrum of pure
H$_2$O \citep{hud93} at the determined temperature was scaled to the
column density derived from the 3.0 and 13 \mum\ bands
(\S\ref{sec:h2o}) and subtracted from each optical depth spectrum. The
shapes and depths of the residual features, and any variations thereof
between sources, are analyzed.  It should be kept in mind, however,
that some features may still be the result of interactions of species
with H$_2$O in solid state matrices, such as the narrow peak induced
in H$_2$O:CO$_2$ mixtures (\S\ref{sec:disc2}; \citealt{kne05, obe07}).

\begin{figure}
\includegraphics[angle=90, scale=0.51]{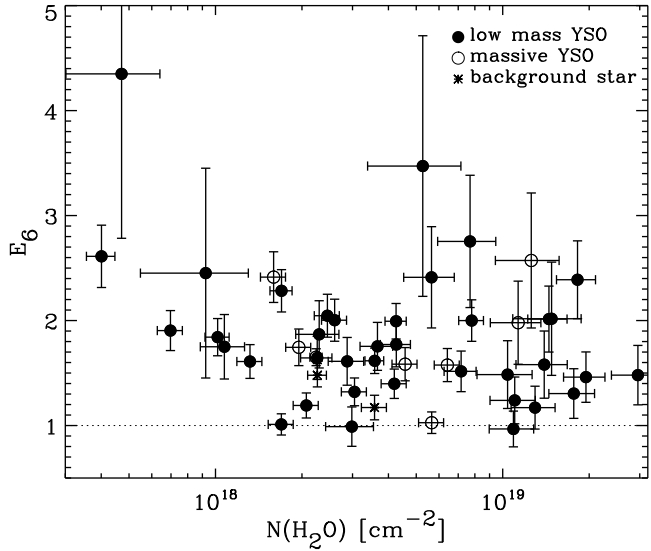}
\caption{Values of $E_6$ (defined in Eq.~\ref{eq:e6}) plotted as a
  function of the solid H$_2$O column density for the entire source
  sample: low mass YSOs (filled circles), massive YSOs (open circles),
  and background stars (stars). For sources near the horizontal dotted
  line the entire 6.0 \mum\ absorption band can be explained by pure
  H$_2$O ice. All other sources have significantly deeper 6.0 \mum\
  bands than expected from the 3 or 13 \mum\ bands.}~\label{f:h2oh2o}
\end{figure}

Toward most low mass YSOs, the observed 6.0 \mum\ absorption band is
deeper than expected from the 3.0 and 13 \mum\ H$_2$O ice bands (Fig.
\ref{f:58}). Such a discrepancy has been reported toward massive YSOs
as well \citep{cox89, sch96, gib00, kea01b} and it is quantified in
units of the H$_2$O column density as follows:

\begin{equation}
E_6=\frac{\int_{1562}^{1785} \tau_{\nu} d\nu}{N({\rm H_2O})\times 1.2\times
10^{-17}\times 0.60} \label{eq:e6}
\end{equation}

\noindent which is the integrated optical depth of the 6.0 \mum\ band
in wavenumber ($\nu$) space divided by the H$_2$O column density
derived from the 3.0 and/or 13 \mum\ bands. The factor of $1.2\times
10^{-17}$ is the integrated band strength of the O-H bending mode in a
pure H$_2$O laboratory ice \citep{ger95} in units of cm/molecule.  The
latter extends all the way out to 8.5 \mum, while the numerator of
Eq.~\ref{eq:e6} integrates only to 6.4 \mum. Therefore the laboratory
band strength is reduced by a factor of 0.60, as derived from the
laboratory spectra.  $E_6$ varies significantly: from $\geq$2 in
$\sim$30\% of the YSOs (e.g.  RNO 91, IRAS 03301+3111, EC 92) to 1.0
(i.e. no excess) in other YSOs (Table~\ref{t:tau} and
Fig.~\ref{f:h2oh2o}).

\begin{figure}
\includegraphics[angle=90, scale=0.43]{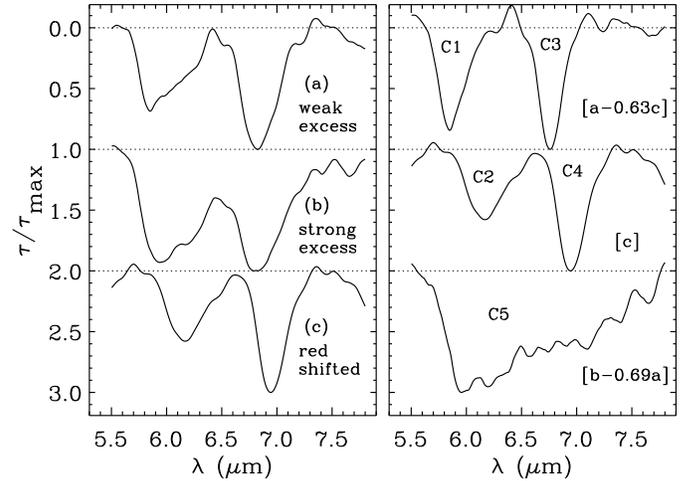}
\caption{Demonstration of the decomposition of the 5-8 \mum\
  absorption complex. In the left panel are shown averaged, smoothed,
  H$_2$O-subtracted spectra of YSOs with small 6.0 \mum\ excess ({\bf
  a}), YSOs with strong 6.0 \mum\ excess ({\bf b}), and the YSO IRAS
  03301+3111 which shows strongly red-shifted 6.0 and 6.85 \mum\ bands
  ({\bf c}).  Applying linear combinations of spectra ({\bf a}), ({\bf
  b}), and ({\bf c}) in the left panel, 5 independent absorption
  components are found and shown in the right panel. Components C1 and
  C3 are found by subtracting spectrum ({\bf c}), multiplied by a
  factor of 0.63, from spectrum ({\bf a}) [{\bf top}]. Components C2
  and C4 are identical to spectrum ({\bf c}) [{\bf middle}].
  Component C5 is found by subtracting spectrum ({\bf a}), multiplied
  by a factor of 0.69, from spectrum ({\bf b}) [{\bf bottom}]. These 5
  components, in addition to the bending mode of H$_2$O, are
  responsible for the 5--8 \mum\ absorption complex in YSOs and
  background stars. Example fits are shown in
  Fig.~\ref{f:compex}. Note that C3 and C4 are the previously found
  short- and long-wavelength components of the 6.85 \mum\ band
  \citep{kea01b}. }  ~\label{f:decomp}
\end{figure}

\begin{figure}
\includegraphics[angle=90, scale=0.57]{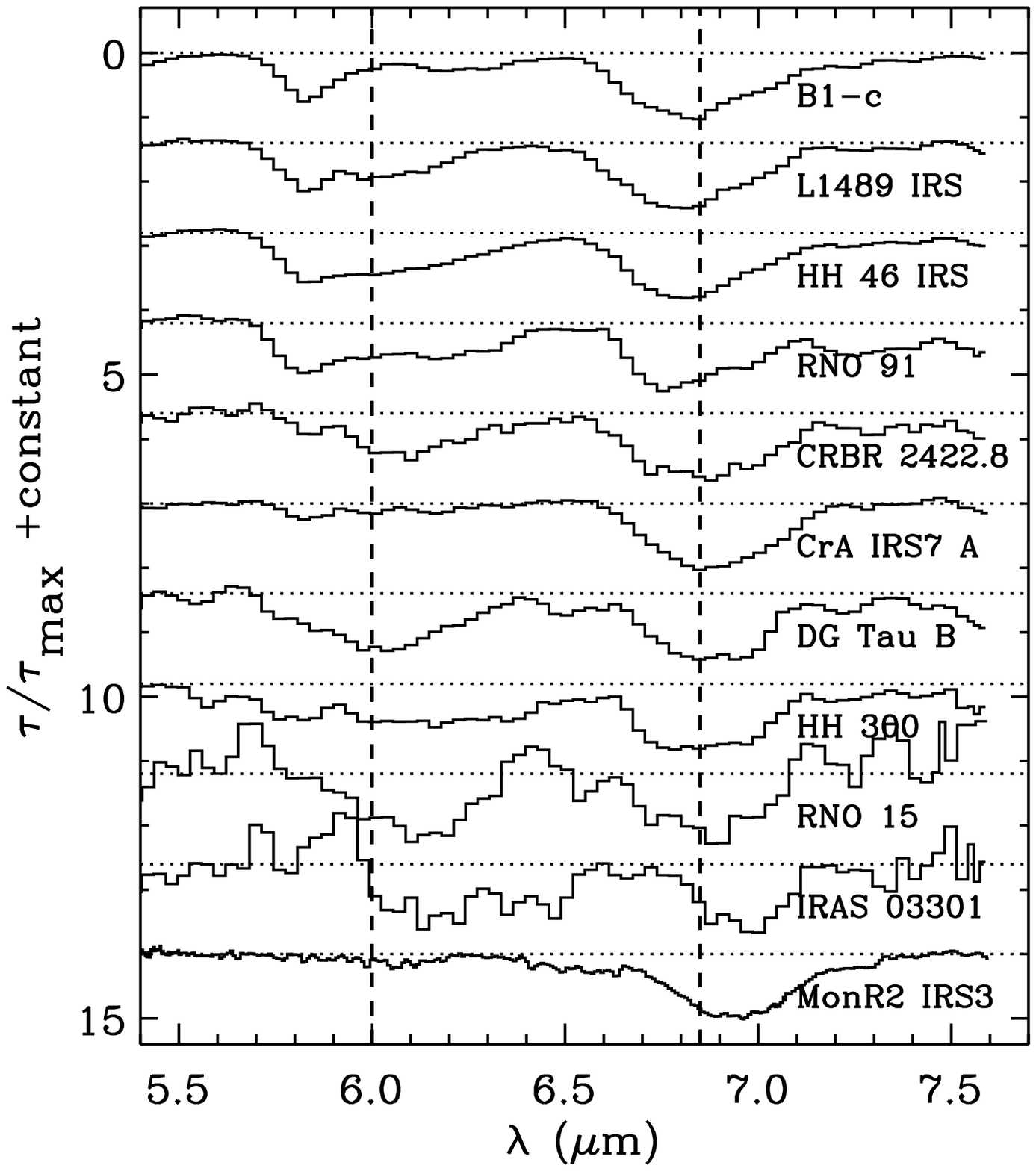}
\caption{Residual absorption in the 5-8 \mum\ region for a selection
  of sources, after subtraction of pure H$_2$O and component C5. The
  vertical dashed lines at 6.0 and 6.85 \mum\ are plotted to
  facilitate comparing the positions of the features in the different
  sources. Note that the plotted sources represent the most extreme
  cases of our sample; most sources resemble the spectrum of HH 46
  IRS.  The spectra are normalized on optical depth scale, and sorted
  from top to bottom in increasing wavelength of the 6 \mum\ residual.
  Note that at the same time the 6.85 \mum\ feature shifts to longer
  wavelengths as well.}  ~\label{f:resid}
\end{figure}

The large source-to-source variations of $E_6$ may hold clues to the
nature of the carrier(s) of the 6.0 \mum\ excess absorption. As is
illustrated in Fig.~\ref{f:decomp}, these variations are used to
separate any independent, overlapping absorption components in the 5-8
\mum\ range.  Initially, a selection of sources with low values of
$E_6$ and sources with high values of $E_6$ is averaged, and the
resulting spectra are scaled and subtracted from each other such that
the distinct peak at 6.85 \mum\ is removed. What remains is a very
broad feature, stretching between 5.8 and 8 \mum.  While the profiles
of the 6.0 and 6.85 \mum\ bands themselves look rather similar for
many sources (e.g. like HH 46 IRS), significant source to source
variations are sometimes observed (Fig.~\ref{f:resid}).  The peak
position of the 6.0 \mum\ component varies between 5.85 and 6.2 \mum,
and broadens as the feature shifts toward longer wavelengths. At the
same time the 6.85 \mum\ band shifts from 6.8 to 6.95 \mum.  By
subtracting extreme spectra from each other, it is found that the
latter consists of two components centered at 6.755 and 6.943 \mum\
with widths of FWHM=0.195 and 0.292 \mum, respectively, similar to
what was previously found for massive protostars \citep{kea01b}.  The
6.0 \mum\ band consists of two components as well, with peak positions
of 5.84 and 6.18 \mum\ and FWHM widths of 0.26 and 0.40 \mum,
respectively.

\begin{figure}
\includegraphics[angle=90, scale=0.57]{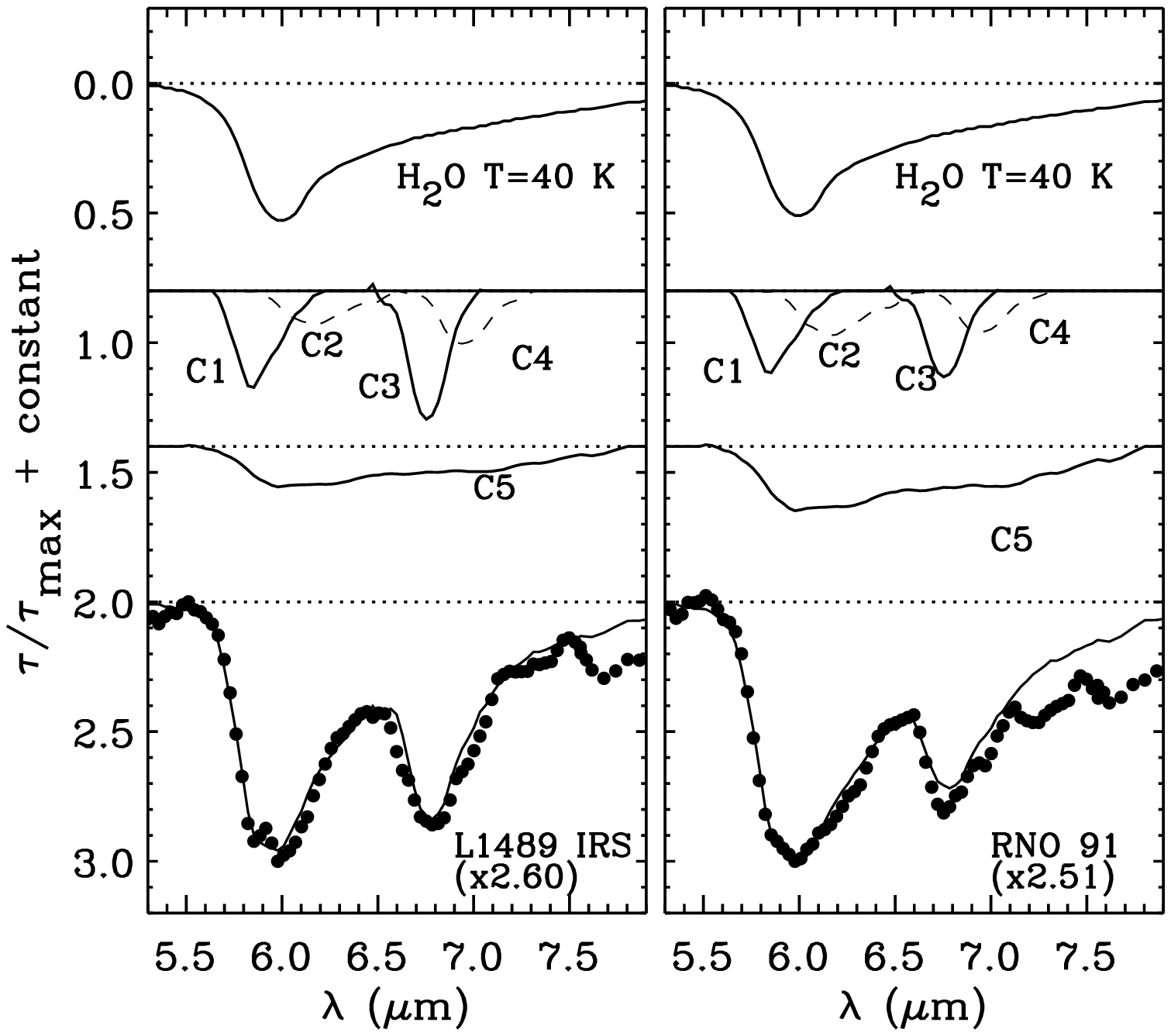}
\caption{Examples of fits to the 5--8 \mum\ absorption complex with
the five components derived in Fig.~\ref{f:decomp}, as well as
H$_2$O. In the left and right panels two different YSOs are shown. The
dots are the observed data points, and the thin line through them is
the sum of all components above.}~\label{f:compex}
\end{figure}

It is thus hypothesized that, after H$_2$O subtraction, the 5--8 \mum\
absorption in all sources is composed of 5 distinct components
(Fig.~\ref{f:decomp}): a set of bands peaking at 5.84 and 6.18 \mum\
(C1 and C2 hereafter), a set of bands peaking at 6.755 and 6.943 \mum\
(C3 and C4), and a broad asymmetric feature stretching between 5.8 and
8 \mum\ and peaking near 5.9 \mum\ (C5). Example fits with these
components are shown in Fig.~\ref{f:compex}.  The peak optical depths
of the components in all sources are listed in Table~\ref{t:tau}.  The
indicated error bars are based on statistical uncertainties in the
spectral data points, except for C5. The accuracy of the C5 optical
depth is affected by the accuracy of the baseline, which at present
can only partly be quantified.  The baseline uncertainty is a
combination of the error in $N$(H$_2$O) (Table~\ref{t:colden}) and the
poorly known error in the continuum choice (\S\ref{sec:cont}). For
many sources the latter, not included in Table~\ref{t:tau}, is likely
at most 0.03 on optical depth scale.

\begin{deluxetable*}{lccccccccc}
\tabletypesize{\scriptsize}
\tablecolumns{10}
\tablewidth{0pc}
\tablecaption{Fit Parameters {\it (publish electronically)}~\label{t:tau}}
\tablehead{
\colhead{Source}& \colhead{$T_{\rm H_2O}  $\tablenotemark{a}}&\colhead{$E_6$\tablenotemark{b}}            & 
                  \colhead{$\tau_{\rm 9.7}$\tablenotemark{c}}&\colhead{$\tau_{\rm 6.0}$\tablenotemark{d}} &
                  \colhead{$\tau_{\rm C1}$\tablenotemark{e}}&\colhead{$\tau_{\rm C2}$\tablenotemark{f}} &
                  \colhead{$\tau_{\rm C3}$\tablenotemark{g}}& \colhead{$\tau_{\rm C4}$\tablenotemark{h}}&
                  \colhead{$\tau_{\rm C5}$\tablenotemark{i}}\\
\colhead{      }& \colhead{               } & \colhead{               } &  
                  \colhead{               } & \colhead{               } & 
                  \colhead{               } & \colhead{               } & 
                  \colhead{               } & \colhead{               } & 
                  \colhead{               } \\}
\startdata
             L1448 IRS1  &      10  &     4.35 (1.57)  &     0.36 (0.04)  &     0.10 (0.01)  &     0.00 (0.01)  &     0.00 (0.01)  &     0.02 (0.01)  &     0.03 (0.01)  &     0.08 (0.01) \\
        IRAS 03235+3004  &      10  &     2.01 (0.31)  &     1.98 (0.20)  &     1.35 (0.07)  &     0.38 (0.15)  &     0.13 (0.15)  &     0.40 (0.07)  &     0.13 (0.07)  &     0.39 (0.06) \\
        IRAS 03245+3002  &      10  &     1.76 (0.25)  &     8.11 (0.81)  &     2.92 (0.17)  &     0.52 (0.40)  &     0.00 (0.40)  &     1.19 (0.16)  &     0.55 (0.16)  &     0.84 (0.13) \\
             L1455 SMM1  &      10  &     2.39 (0.37)  &     5.18 (0.52)  &     1.92 (0.08)  &     0.55 (0.24)  &     0.17 (0.24)  &     0.86 (0.16)  &     0.25 (0.16)  &     0.83 (0.07) \\
                 RNO 15  &  10$^*$  &     1.90 (0.19)  & emission         &     0.08 (0.01)  &     0.01 (0.01)  &     0.04 (0.01)  &     0.03 (0.01)  &     0.03 (0.01)  &     0.02 (0.01) \\
             L1455 IRS3  &      10  &     2.45 (1.00)  &     0.15 (0.05)  &     0.13 (0.01)  &     0.08 (0.04)  &     0.08 (0.04)  &     0.04 (0.03)  &     0.02 (0.03)  &     0.00 (0.03) \\
        IRAS 03254+3050  &  40$^*$  &     1.75 (0.23)  &     1.63 (0.16)  &     0.32 (0.02)  &     0.14 (0.06)  &     0.04 (0.06)  &     0.18 (0.02)  &     0.09 (0.02)  &     0.04 (0.01) \\
        IRAS 03271+3013  &      10  &     2.75 (0.63)  &     1.29 (0.13)  &     0.95 (0.06)  &     0.39 (0.11)  &     0.16 (0.11)  &     0.33 (0.13)  &     0.07 (0.13)  &     0.26 (0.04) \\
        IRAS 03301+3111  &  40$^*$  &     2.61 (0.30)  &     0.49 (0.05)  &     0.06 (0.01)  &     0.00 (0.00)  &     0.03 (0.00)  &     0.00 (0.01)  &     0.03 (0.01)  &     0.02 (0.01) \\
                   B1-a  &      10  &     1.49 (0.32)  &     2.35 (0.23)  &     0.73 (0.04)  &     0.33 (0.13)  &     0.12 (0.13)  &     0.39 (0.04)  &     0.29 (0.04)  &     0.03 (0.05) \\
                   B1-c  &      10  &     1.48 (0.28)  &     7.03 (0.70)  &     2.04 (0.08)  &     0.65 (0.37)  &     0.14 (0.37)  &     1.01 (0.06)  &     0.64 (0.06)  &     0.24 (0.13) \\
                   B1-b  &      10  &     1.30 (0.24)  &     2.30 (0.23)  &     1.15 (0.04)  &     0.51 (0.21)  &     0.05 (0.21)  &     0.52 (0.04)  &     0.35 (0.04)  &     0.00 (0.07) \\
                B5 IRS3  &  10$^*$  &     1.84 (0.18)  &     0.32 (0.04)  &     0.12 (0.02)  &     0.01 (0.01)  &     0.04 (0.01)  &     0.06 (0.02)  &     0.05 (0.02)  &     0.02 (0.02) \\
                B5 IRS1  &  10$^*$  &     1.64 (0.21)  &     1.23 (0.12)  &     0.19 (0.01)  &     0.07 (0.01)  &     0.03 (0.01)  &     0.07 (0.02)  &     0.03 (0.02)  &     0.01 (0.01) \\
              L1489 IRS  &  40$^*$  &     1.77 (0.21)  &     1.32 (0.13)  &     0.37 (0.02)  &     0.15 (0.06)  &     0.05 (0.06)  &     0.19 (0.02)  &     0.08 (0.02)  &     0.06 (0.02) \\
       IRAS 04108+2803B  &  40$^*$  &     1.61 (0.23)  &     0.49 (0.05)  &     0.22 (0.01)  &     0.10 (0.03)  &     0.06 (0.03)  &     0.10 (0.01)  &     0.06 (0.01)  &     0.01 (0.02) \\
                 HH 300  &  10$^*$  &     2.00 (0.20)  &     1.09 (0.11)  &     0.24 (0.01)  &     0.04 (0.03)  &     0.05 (0.03)  &     0.08 (0.02)  &     0.06 (0.02)  &     0.04 (0.01) \\
               DG Tau B  & 100$^*$  &     1.87 (0.32)  &     2.05 (0.20)  &     0.23 (0.02)  &     0.09 (0.02)  &     0.08 (0.02)  &     0.08 (0.02)  &     0.10 (0.02)  &     0.01 (0.01) \\
              HH 46 IRS  &      40  &     2.00 (0.20)  &     2.23 (0.22)  &     0.75 (0.04)  &     0.34 (0.13)  &     0.17 (0.13)  &     0.36 (0.02)  &     0.18 (0.02)  &     0.10 (0.02) \\
        IRAS 12553-7651  &      10  &     0.99 (0.19)  &     1.40 (0.14)  &     0.15 (0.01)  &     0.01 (0.03)  &     0.02 (0.03)  &     0.09 (0.03)  &     0.07 (0.03)  &     0.00 (0.02) \\
        IRAS 13546-3941  &      10  &     1.19 (0.12)  &     0.23 (0.03)  &     0.15 (0.01)  &     0.05 (0.03)  &     0.03 (0.03)  &     0.05 (0.01)  &     0.02 (0.01)  &     0.00 (0.01) \\
        IRAS 15398-3359  &      10  &     2.02 (0.54)  &     3.32 (0.33)  &     1.41 (0.08)  &     0.54 (0.17)  &     0.15 (0.17)  &     0.55 (0.03)  &     0.31 (0.03)  &     0.40 (0.09) \\
               Elias 29  &  40$^*$  &     1.32 (0.13)  &     1.44 (0.14)  &     0.24 (0.02)  &     0.05 (0.02)  &     0.04 (0.02)  &     0.05 (0.02)  &     0.02 (0.02)  &     0.00 (0.02) \\
        CRBR 2422.8-342  &  40$^*$  &     1.40 (0.14)  &     1.68 (0.17)  &     0.32 (0.03)  &     0.06 (0.05)  &     0.09 (0.05)  &     0.13 (0.02)  &     0.11 (0.02)  &     0.00 (0.02) \\
                 RNO 91  &  40$^*$  &     1.99 (0.17)  &     0.80 (0.08)  &     0.39 (0.02)  &     0.13 (0.05)  &     0.07 (0.05)  &     0.13 (0.01)  &     0.06 (0.01)  &     0.10 (0.02) \\
        IRAS 17081-2721  &  10$^*$  &     1.61 (0.16)  & emission         &     0.10 (0.01)  &     0.04 (0.02)  &     0.01 (0.02)  &     0.07 (0.01)  &     0.03 (0.01)  &     0.01 (0.01) \\
 SSTc2dJ171122.2-272602  &      10  &     1.58 (0.32)  &     3.09 (0.31)  &     1.00 (0.05)  &     0.24 (0.14)  &     0.13 (0.14)  &     0.48 (0.04)  &     0.30 (0.04)  &     0.19 (0.06) \\
 2MASSJ17112317-2724315  &      10  &     1.46 (0.24)  &     4.58 (0.46)  &     1.30 (0.06)  &     0.35 (0.24)  &     0.04 (0.24)  &     0.74 (0.04)  &     0.42 (0.04)  &     0.37 (0.07) \\
                  EC 74  &  10$^*$  &     1.75 (0.31)  &     0.47 (0.05)  &     0.10 (0.01)  &     0.02 (0.03)  &     0.03 (0.03)  &     0.05 (0.02)  &     0.02 (0.02)  &     0.01 (0.01) \\
                  EC 82  &      10  &  $<$0.63         & emission         &     0.04 (0.01)  &     0.02 (0.02)  &     0.00 (0.02)  &     0.02 (0.01)  &     0.01 (0.01)  &     0.00 (0.01) \\
                SVS 4-5  &      10  &     2.41 (0.48)  &     1.61 (0.16)  &     0.68 (0.04)  &     0.31 (0.11)  &     0.09 (0.11)  &     0.28 (0.02)  &     0.21 (0.02)  &     0.17 (0.03) \\
                  EC 90  &      10  &     1.01 (0.10)  & emission         &     0.13 (0.01)  &     0.05 (0.02)  &     0.02 (0.02)  &     0.06 (0.01)  &     0.03 (0.01)  &     0.00 (0.01) \\
                  EC 92  &  40$^*$  &     2.28 (0.20)  & emission         &     0.24 (0.02)  &     0.15 (0.04)  &     0.07 (0.04)  &     0.10 (0.02)  &     0.02 (0.02)  &     0.02 (0.01) \\
                    CK4  &      10  &  $<$1.08         & emission         &     0.08 (0.00)  &     0.03 (0.03)  &     0.04 (0.03)  &     0.01 (0.02)  &     0.01 (0.02)  &     0.00 (0.02) \\
            R CrA IRS 5  &  40$^*$  &     1.62 (0.12)  &     1.02 (0.10)  &     0.29 (0.02)  &     0.11 (0.06)  &     0.09 (0.06)  &     0.13 (0.01)  &     0.08 (0.01)  &     0.00 (0.01) \\
             HH 100 IRS  &  10$^*$  &     2.05 (0.20)  &     0.69 (0.07)  &     0.27 (0.03)  &     0.10 (0.05)  &     0.10 (0.05)  &     0.11 (0.03)  &     0.09 (0.03)  &     0.00 (0.02) \\
             CrA IRS7 A  &      10  &     0.97 (0.17)  &     3.02 (0.30)  &     0.54 (0.04)  &     0.08 (0.09)  &     0.03 (0.09)  &     0.26 (0.02)  &     0.35 (0.02)  &     0.00 (0.04) \\
             CrA IRS7 B  &      10  &     1.24 (0.22)  &     3.44 (0.34)  &     0.68 (0.04)  &     0.19 (0.12)  &     0.06 (0.12)  &     0.35 (0.02)  &     0.31 (0.02)  &     0.04 (0.04) \\
             CrA IRAS32  &      10  &     3.47 (1.24)  &     2.43 (0.25)  &     0.86 (0.04)  &     0.59 (0.29)  &     0.41 (0.29)  &     0.34 (0.22)  &     0.31 (0.22)  &     0.00 (0.05) \\
              L1014 IRS  &      10  &     1.52 (0.19)  &     2.08 (0.21)  &     0.62 (0.04)  &     0.36 (0.15)  &     0.21 (0.15)  &     0.34 (0.04)  &     0.16 (0.04)  &     0.00 (0.02) \\
        IRAS 23238+7401  &      10  &     1.17 (0.20)  &     2.30 (0.23)  &     0.72 (0.03)  &     0.29 (0.10)  &     0.07 (0.10)  &     0.38 (0.05)  &     0.16 (0.05)  &     0.00 (0.05) \\
                W3 IRS5  &  10$^*$  &     1.03 (0.10)  &     6.28 (0.63)  &     0.29 (0.01)  &     0.00 (0.04)  &     0.00 (0.04)  &     0.14 (0.01)  &     0.16 (0.01)  &     0.00 (0.02) \\
             MonR2 IRS3  & 100$^*$  &     2.41 (0.24)  &     2.25 (0.23)  &     0.20 (0.01)  &     0.01 (0.01)  &     0.02 (0.01)  &     0.06 (0.02)  &     0.22 (0.02)  &     0.09 (0.01) \\
                  GL989  &  40$^*$  &     1.65 (0.08)  &     0.66 (0.07)  &     0.18 (0.01)  &     0.06 (0.03)  &     0.04 (0.03)  &     0.10 (0.01)  &     0.07 (0.01)  &     0.01 (0.01) \\
                   W33A  &      40  &     2.57 (0.64)  &     6.02 (0.60)  &     1.87 (0.03)  &     0.46 (0.20)  &     0.28 (0.20)  &     0.74 (0.05)  &     0.51 (0.05)  &     0.67 (0.07) \\
                GL7009S  &      10  &     1.98 (0.40)  &     3.62 (0.36)  &     1.17 (0.05)  &     0.38 (0.18)  &     0.01 (0.18)  &     0.72 (0.04)  &     0.54 (0.04)  &     0.37 (0.06) \\
                 GL2136  & 100$^*$  &     1.58 (0.16)  &     3.02 (0.30)  &     0.35 (0.01)  &     0.11 (0.05)  &     0.04 (0.05)  &     0.19 (0.02)  &     0.14 (0.02)  &     0.06 (0.01) \\
              S140 IRS1  & 100$^*$  &     1.74 (0.17)  &     1.93 (0.19)  &     0.16 (0.01)  &     0.05 (0.02)  &     0.03 (0.02)  &     0.05 (0.01)  &     0.10 (0.01)  &     0.00 (0.01) \\
           NGC7538 IRS9  &  10$^*$  &     1.58 (0.16)  &     2.25 (0.23)  &     0.48 (0.02)  &     0.12 (0.08)  &     0.07 (0.08)  &     0.27 (0.01)  &     0.15 (0.01)  &     0.04 (0.02) \\
               Elias 16  &  10$^*$  &     1.48 (0.11)  &     0.73 (0.07)  &     0.18 (0.01)  &     0.04 (0.02)  &     0.04 (0.02)  &     0.07 (0.01)  &     0.03 (0.01)  &     0.00 (0.02) \\
                 EC 118  &      10  &     1.17 (0.12)  &     1.42 (0.14)  &     0.23 (0.02)  &     0.06 (0.05)  &     0.04 (0.05)  &     0.20 (0.03)  &     0.10 (0.03)  &     0.00 (0.01) \\
\enddata
\tablecomments{Uncertainties in parentheses based on statistical errors in the spectra only,
 unless noted otherwise below.}
\tablenotetext{a}{ Temperature of pure H$_2$O laboratory ice \citep{hud93} assumed. Values indicated with an asterisk are most accurate because of the availability of good quality L-band spectra.}
\tablenotetext{b}{ 6.0 \mum\ excess as defined in Eq.~\ref{eq:e6}. Uncertainty includes uncertainty in $N$(H$_2$O).}
\tablenotetext{c}{ peak optical depth at 9.7 \mum, without correction for underlying emission. Uncertainty includes 10\% of $\tau_{9.7}$ to account for errors in the continuum determination.}
\tablenotetext{d}{ peak optical depth at 6.0 \mum, including H$_2$O absorption.}
\tablenotetext{e}{ peak optical depth component C1}
\tablenotetext{f}{ peak optical depth component C2 }
\tablenotetext{g}{ peak optical depth component C3 }
\tablenotetext{h}{ peak optical depth component C4 }
\tablenotetext{i}{ peak optical depth component C5. Uncertainty includes uncertainty in $N$(H$_2$O) as this affects the baseline level. The continuum uncertainty is not included. It is likely on the order of 0.03 for most sources.}
\end{deluxetable*}

\begin{figure}
\includegraphics[angle=90, scale=0.52]{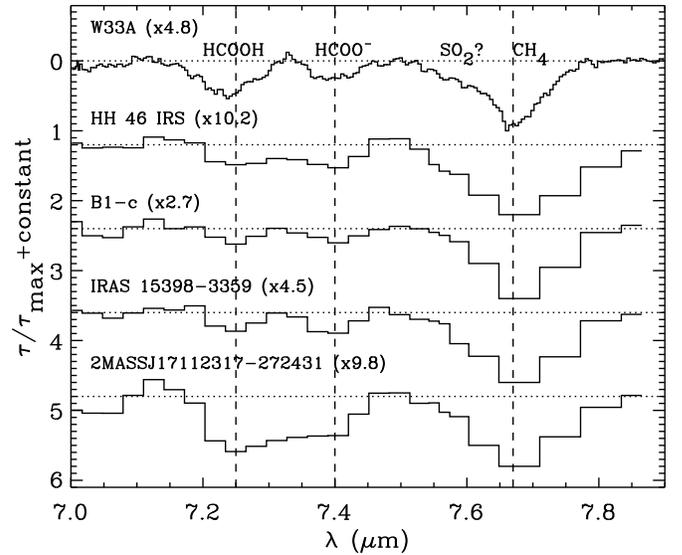}
\caption{Optical depth 7--8 \mum\ Spitzer/IRS spectra of sources
  showing the 7.25, 7.40, and 7.67 \mum\ features. The top source,
  W33A, is a massive YSO of which a higher resolution ($R$=800)
  ISO/SWS spectrum is shown. The features are labeled with their
  (possible) identifications.}  ~\label{f:7_8um}
\end{figure}

\subsubsection{Weak Features in the 7--8 $\mu m$ Wavelength Range}~\label{sec:7_8}

In addition to solid H$_2$O and the 5 components discussed above, a
number of weak features can be discerned in the 7--8 \mum\ region of
several low mass YSOs, much like those identified toward massive
protostars \citep{sch99}. These features lie on top of a curved
`continuum' created by the wings of the prominent 6 and 6.85 \mum\
features.  A high order polynomial (up to 6$^{\rm th}$) had to be used
to fit a local baseline to typically these wavelength regions:
7.00--7.14, 7.3--7.35, 7.47--7.50, and 7.80--7.95 \mum.  Examples of
the resulting features are shown in Fig.~\ref{f:7_8um}, where they are
compared to the higher resolution ($R=800$) ISO/SWS spectrum of the
massive YSO W33A. The 7.25 \mum\ feature is likely due to the HCOOH
molecule (\S\ref{sec:hcooh}).  The feature at 7.40 \mum\ is possibly
due to HCOO$^-$ (\S\ref{sec:hcoom} ;\citealt{sch99}), while the 7.67
\mum\ feature is identified with CH$_4$ \citep{boo96, obe08}.

\subsubsection{Substructures in the 9.7 \mum\ Silicates Absorption Band}~\label{sec:nh3ch3oh}

\begin{figure}
\includegraphics[angle=90, scale=0.42]{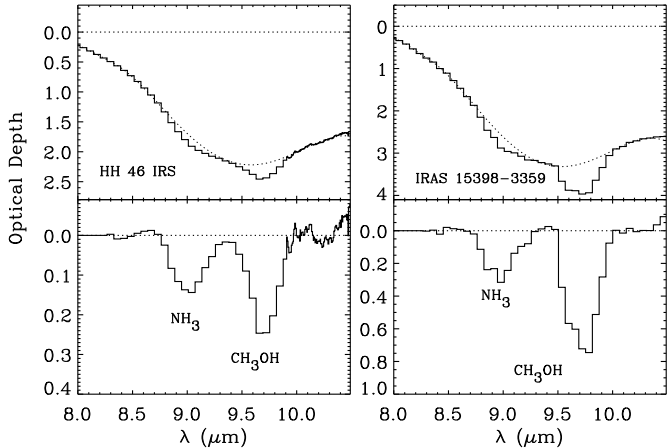}
\caption{Features of solid NH$_3$ and CH$_3$OH in the spectra of the
low mass YSOs HH~46 IRS (left panels) and IRAS 15398-3359 (right). In
the top panels for each source the dashed line indicates a local
polynomial continuum, which was used to derive the optical depth plots
in the lower panels.}  ~\label{f:8_10um}
\end{figure}

Several sources show substructures at 9.0 and 9.7 \mum\ within the
strong silicate absorption band (Fig.~\ref{f:8_10um}).  Previous work
indicated that these can be attributed to the umbrella mode of solid
NH$_3$ and the C-O stretching mode of CH$_3$OH (e.g. \citealt{lac98,
  ski92}).  To extract these and any other ice absorption features in
the 8-12 \mum\ range, the silicate absorption and possible blended
emission needs to be accurately modeled. This is deferred to a future
publication (S. Bottinelli, in prep.), and here a local 4$^{\rm th}$
order polynomial continuum is fitted to the wavelength regions
8.25-8.75, 9.23-9.37, and 9.98-10.4 \mum\ to extract the features
(Fig.~\ref{f:8_10um}). Column densities of CH$_3$OH and NH$_3$ derived
from these and other features are further discussed in
\S\S\ref{sec:ch3oh} and~\ref{sec:nh3}.

\section{Constraining the Origin of the 5--8 \mum\ Absorption Features}~\label{sec:id}

Numerous carriers have been proposed to account for the various
features in the 5--8 \mum\ wavelength region \citep{tie84, kea01b}.
Here new observational constraints to possible carriers are discussed,
using the present large source sample and wavelength coverage. First,
the contributions of the most securely identified (CH$_3$OH, NH$_3$)
or otherwise well constrained species (H$_2$CO, HCOOH, HCOO$^-$) to
the observed C1-C5 components are discussed
(\S\ref{sec:carr}). Subsequently, the origin of the remaining
absorption is constrained by correlating it with a number of other
observables (\S\ref{sec:corr}).

\subsection{Absorption by Known Species}~\label{sec:carr}

\subsubsection{CH$_3$OH}~\label{sec:ch3oh}

\begin{figure}
\includegraphics[angle=90, scale=0.41]{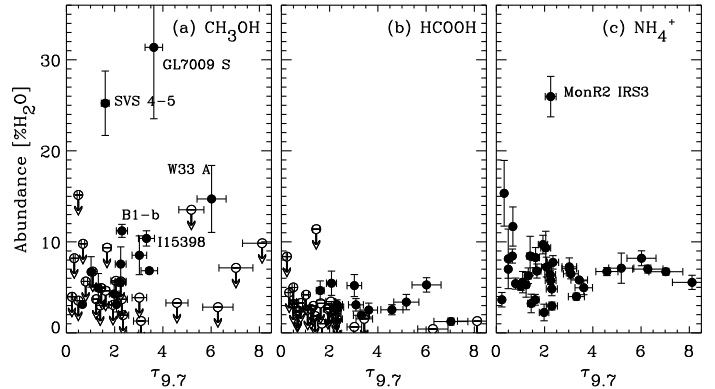}
\caption{Abundances of solid CH$_3$OH ({\bf panel a}), HCOOH ({\bf
  b}), and NH$_4^+$ ({\bf c}) in percentage of H$_2$O as a function of
  the 9.7 \mum\ silicate band optical depth. The solid CH$_3$OH
  abundance is clearly more sight-line dependent than the HCOOH
  abundance.  Sources with the largest relative CH$_3$OH column
  densities are labeled in the left panel. Open symbols with arrows
  refer to 3$\sigma$ upper limits.  Error bars are 1$\sigma$. The
  NH$_4^+$ abundance in the right panel was calculated assuming both
  the C3 and C4 components are due to NH$_4^+$, but after subtracting
  the contributions from CH$_3$OH and H$_2$CO.}  ~\label{f:abun}
\end{figure}

Solid CH$_3$OH was previously unambiguously detected by its C$-$H
stretching mode at 3.54 \mum\ (e.g. \citealt{gri91}) and its C$-$O
stretching mode at 9.7 \mum\ \citep{ski92}.  In our sample, a Gaussian
was fitted to the 9.7 \mum\ feature on an optical depth scale and a
CH$_3$OH detection is claimed if the integrated optical depth has a
larger than 3$\sigma$ significance and if the peak position and FWHM
are similar to previous detections, i.e.  within 9.68-9.76 and
0.25-0.36 \mum\ respectively.  The 3.54 \mum\ band, when available,
was treated similarly.  Thus, CH$_3$OH was positively identified in 12
low mass YSOs (Table~\ref{t:colden}; Fig.~\ref{f:abun}). Of those, 6
are based on the 9.7 \mum\ band only.  Column densities were
calculated using integrated band strengths of 5.6$\times 10^{-18}$ and
1.6$\times 10^{-17}$ cm/molecule for the 3.54 and 9.7 \mum\ bands,
respectively \citep{ker99}.  In agreement with previous studies
\citep{dar99, pon03b}, the CH$_3$OH column varies considerably with
respect to solid H$_2$O: from low upper limits of a few percent
(3$\sigma$) in several sources, to detections of up to 30\% in a few
others.  In the 5--8 \mum\ region, the C$-$H deformation mode of
CH$_3$OH overlaps with the C3 component, and, at the derived column
densities, contributes typically 5-20\% to its strength, but up to
40\% for SVS 4-5 and GL7009S.

\begin{deluxetable*}{lrrrr}
\tabletypesize{\scriptsize}
\tablecolumns{5}
\tablewidth{0pc}
\tablecaption{Column Densities of the Ices~\label{t:colden}}
\tablehead{
\colhead{Source}& \colhead{$N$(H$_2$O)}     & \colhead{$N$(HCOOH)\tablenotemark{c}}      & 
                  \colhead{$N$(CH$_3$OH)}   & \colhead{$N$(NH$_4^+$)\tablenotemark{e}} \\
\colhead{      }& \colhead{$10^{18}$ \sqcm} & \colhead{\% H$_2$O      } & 
                  \colhead{\% H$_2$O      } & \colhead{\% H$_2$O      } \\}
\startdata
             L1448 IRS1            &     0.47          (0.16)            &        $\leq$12.1          &         $<$14.9                                      &     13.9  (4.6) \\
        IRAS 03235+3004            &    14.48   (2.26)\tablenotemark{b}  &         $\leq$2.7          &                       4.2           (1.2)            &      2.2  (0.9) \\
        IRAS 03245+3002            &    39.31   (5.65)\tablenotemark{b}  &         $\leq$1.2          &          $<$9.8                                      &      5.5  (0.7) \\
             L1455 SMM1            &    18.21   (2.82)\tablenotemark{b}  &               3.3   (0.8)  &         $<$13.5                                      &      7.1  (1.6) \\
                 RNO 15            &     0.69          (0.06)            &         $\leq$5.8          &   $<$5.0\tablenotemark{d}                            &     13.5  (2.4) \\
             L1455 IRS3            &     0.92   (0.37)\tablenotemark{b}  &        $\leq$15.1          &         $<$12.5                                      &      8.9  (5.5) \\
        IRAS 03254+3050            &     3.66          (0.47)            &         $\leq$3.2          &          $<$4.6                                      &      8.2  (1.1) \\
        IRAS 03271+3013            &     7.69   (1.76)\tablenotemark{b}  &         $\leq$2.5          &          $<$5.6                                      &      5.8  (3.3) \\
        IRAS 03301+3111            &     0.40          (0.04)            &         $\leq$4.9          &         $<$15.1                                      &      8.1  (3.2) \\
                   B1-a            &    10.39   (2.26)\tablenotemark{b}  &            $<$2.8          &          $<$1.9                                      &      7.7  (0.7) \\
                   B1-c            &    29.55   (5.65)\tablenotemark{b}  &               1.2   (0.3)  &          $<$7.1                                      &      6.7  (0.3) \\
                   B1-b            &    17.67   (3.20)\tablenotemark{b}  &               3.1   (0.5)  &                      11.2           (0.7)            &      2.9  (0.4) \\
                B5 IRS3            &     1.01          (0.09)            &         $\leq$4.4          &          $<$8.1                                      &     15.3  (3.6) \\
         B5 IRS1\tablenotemark{a}  &     2.26          (0.28)            &         $\leq$2.6          &          $<$3.7                                      &      5.2  (1.3) \\
              L1489 IRS            &     4.26          (0.51)            &         $\leq$2.9          &                       4.9    (1.5)\tablenotemark{d}  &      6.2  (0.8) \\
       IRAS 04108+2803B            &     2.87          (0.40)            &         $\leq$3.1          &          $<$3.5                                      &      6.9  (0.9) \\
                 HH 300            &     2.59          (0.25)            &         $\leq$2.5          &          $<$6.7                                      &      5.5  (1.1) \\
               DG Tau B            &     2.29          (0.39)            &         $\leq$3.4          &          $<$5.7                                      &      9.3  (1.8) \\
       HH 46 IRS\tablenotemark{a}  &     7.79   (0.77)\tablenotemark{b}  &               2.7   (0.7)  &                       5.5    (0.3)\tablenotemark{d}  &      6.3  (0.4) \\
        IRAS 12553-7651            &     2.98   (0.56)\tablenotemark{b}  &         $\leq$1.8          &          $<$3.0                                      &      8.4  (2.1) \\
        IRAS 13546-3941            &     2.07   (0.20)\tablenotemark{b}  &            $<$8.3          &          $<$3.9                                      &      3.6  (0.7) \\
        IRAS 15398-3359            &    14.79   (3.95)\tablenotemark{b}  &               1.9   (0.3)  &                      10.3           (0.8)            &      3.9  (0.3) \\
        Elias 29\tablenotemark{a}  &     3.04          (0.30)            &        $\leq$11.3          &   $<$4.9\tablenotemark{d}                            &      3.2  (1.0) \\
 CRBR 2422.8-342\tablenotemark{a}  &     4.19          (0.41)            &         $\leq$1.0          &          $<$9.3                                      &      6.7  (0.7) \\
                 RNO 91            &     4.25          (0.36)            &         $\leq$3.2          &          $<$5.6                                      &      5.4  (0.6) \\
        IRAS 17081-2721            &     1.31          (0.13)            &         $\leq$2.4          &                       3.3    (0.8)\tablenotemark{d}  &      9.9  (1.8) \\
 SSTc2dJ171122.2-272602            &    13.94   (2.82)\tablenotemark{b}  &               3.0   (0.6)  &          $<$1.3                                      &      6.4  (0.5) \\
 2MASSJ17112317-2724315            &    19.49   (3.20)\tablenotemark{b}  &               2.5   (0.5)  &          $<$3.2                                      &      6.7  (0.4) \\
                  EC 74            &     1.07          (0.18)            &         $\leq$3.2          &   $<$9.3\tablenotemark{d}                            &      9.7  (2.8) \\
           EC 82\tablenotemark{a}  &     0.39          (0.07)            &         $\leq$2.5          &         $<$14.2                                      &     10.2  (6.3) \\
         SVS 4-5\tablenotemark{a}  &     5.65          (1.13)            &               4.6   (1.0)  &                      25.2    (3.5)\tablenotemark{d}  &      3.6  (0.5) \\
           EC 90\tablenotemark{a}  &     1.69          (0.16)            &         $\leq$4.1          &                       6.8           (1.6)            &      4.4  (0.9) \\
           EC 92\tablenotemark{a}  &     1.69          (0.14)            &         $\leq$5.3          &                      11.7    (3.5)\tablenotemark{d}  &      2.9  (1.8) \\
                    CK4            &  $<$1.56                            &            \ldots          &                    \ldots                            &   \ldots        \\
            R CrA IRS 5            &     3.58          (0.26)            &         $\leq$4.1          &                       6.6    (1.6)\tablenotemark{d}  &      5.1  (0.4) \\
      HH 100 IRS\tablenotemark{a}  &     2.45          (0.24)            &         $\leq$2.4          &   $<$9.7\tablenotemark{d}                            &     11.6  (2.1) \\
             CrA IRS7 A            &    10.89   (1.92)\tablenotemark{b}  &         $\leq$0.6          &          $<$3.8                                      &      7.0  (0.4) \\
             CrA IRS7 B            &    11.01   (1.97)\tablenotemark{b}  &         $\leq$1.5          &                       6.8           (0.3)            &      5.8  (0.2) \\
             CrA IRAS32            &     5.26   (1.88)\tablenotemark{b}  &         $\leq$9.8          &         $<$18.1                                      &     15.9  (7.9) \\
              L1014 IRS            &     7.16   (0.91)\tablenotemark{b}  &               5.4   (1.3)  &                       3.1           (0.8)            &      7.2  (1.0) \\
        IRAS 23238+7401            &    12.95   (2.26)\tablenotemark{b}  &         $\leq$1.5          &          $<$3.6                                      &      4.8  (0.7) \\
         W3 IRS5\tablenotemark{a}  &     5.65          (0.56)            &         $\leq$0.4          &   $<$2.8\tablenotemark{d}                            &      7.0  (0.4) \\
      MonR2 IRS3\tablenotemark{a}  &     1.59          (0.15)            &         $\leq$2.7          &   $<$5.6\tablenotemark{d}                            &     25.9  (2.2) \\
           GL989\tablenotemark{a}  &     2.24          (0.10)            &         $\leq$2.2          &                       3.1           (0.1)            &      8.4  (0.7) \\
            W33A\tablenotemark{a}  &    12.57          (3.14)            &               5.2   (0.7)  &                      14.7    (3.6)\tablenotemark{d}  &      8.1  (0.8) \\
         GL7009S\tablenotemark{a}  &    11.31          (2.26)            &               2.5   (0.7)  &                      31.3    (7.8)\tablenotemark{d}  &      4.9  (0.7) \\
          GL2136\tablenotemark{a}  &     4.57          (0.45)            &               5.1   (1.2)  &                       8.5    (2.1)\tablenotemark{d}  &      7.2  (0.9) \\
       S140 IRS1\tablenotemark{a}  &     1.95          (0.19)            &            $<$2.6          &   $<$3.0\tablenotemark{d}                            &      9.6  (1.0) \\
    NGC7538 IRS9\tablenotemark{a}  &     6.41          (0.64)            &            $<$2.0          &                       7.5    (1.8)\tablenotemark{d}  &      5.7  (0.4) \\
        Elias 16\tablenotemark{a}  &     2.26          (0.16)            &         $\leq$1.4          &          $<$2.3                                      &      5.2  (0.5) \\
          EC 118\tablenotemark{a}  &     3.57          (0.35)            &         $\leq$1.4          &          $<$4.7                                      &     11.6  (1.4) \\
\enddata
\tablecomments{Uncertainties (1 $\sigma$) are indicated in brackets and upper limits are of 3 $\sigma$ significance.}
\tablenotetext{a}{Value obtained from or comparable to previous work
or references therein:
\citealt{dar99} (CH$_3$OH: W33A, GL2136, W3 IRS5, S140 IRS1, NGC7538
IRS9, MonR2 IRS3, HH 100 IRS, GL7009S),
\citealt{boo00_2} (CH$_3$OH: Elias 29), 
\citealt{boo04_2} (H$_2$O and CH$_3$OH: B5 IRS1 and HH 46 IRS),
\citealt{bro99} (H$_2$O: EC 90), 
\citealt{eir89} (H$_2$O: EC 82), 
\citealt{kea01b} (H$_2$O: W33A, GL2136, W3 IRS5, S140 IRS1, NGC7538
IRS9, MonR2 IRS3, HH 100 IRS, GL989, GL7009S),
\citealt{kne05} (H$_2$O and CH$_3$OH: Elias 16 and EC 118),
\citealt{pon04} (H$_2$O and  CH$_3$OH in EC 92, SVS 4-5), 
\citealt{pon05} (H$_2$O and CH$_3$OH: CRBR 2422.8-3423).}
\tablenotetext{b}{13 \mum\ H$_2$O libration mode used for $N$(H$_2$O) determination (3 \mum\ band in other cases).}
\tablenotetext{c}{Detections and 3-$\sigma$ upper limits of $N$(HCOOH) based on 7.25 \mum\ C-H deformation mode. Sources for which the 5.85 \mum\ C=O stretching mode provides tighter constraints are indicated with '$\leq$'.}
\tablenotetext{d}{The 3.53\ \mum\ band provides a better or comparable constraint to $N$(CH$_3$OH) than does the 9.7 \mum\ band.}
\tablenotetext{e}{Assuming both C3 and C4 components are due to NH$_4^+$, which is still a matter of debate (\S\ref{sec:disc3c4}); integrated band strength $A=4.4\times 10^{-17}$ cm/molecule assumed \citep{sch03}.}
\end{deluxetable*}

\subsubsection{HCOOH}~\label{sec:hcooh}

\begin{figure}[t]
\includegraphics[angle=90, scale=0.44]{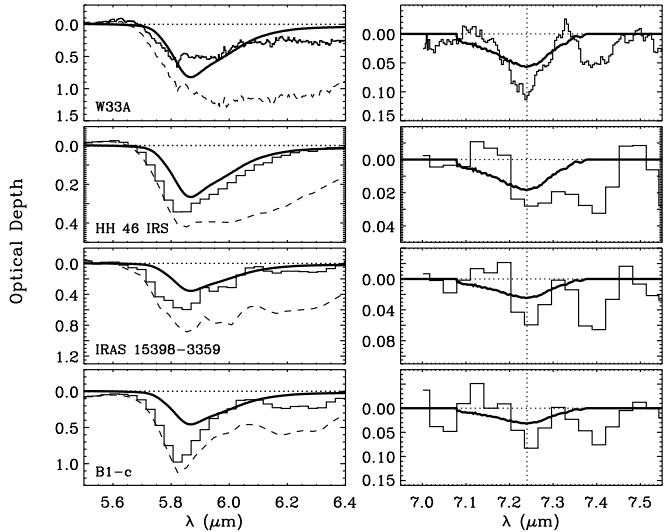}
\caption{Comparison of laboratory solid HCOOH and interstellar
  spectra. The left panels show the region of the C$=$O stretching
  mode, the right panels the C-H deformation mode.  From top to bottom
  four YSOs are shown with confirmed 7.25 \mum\ absorption band
  detections (indicated with vertical dotted lines in the right
  panels).  The histograms in the left panels represent component C1
  of the interstellar spectra, i.e. after subtracting solid H$_2$O and
  components C2 and C5 defined in \S\ref{sec:60}.  The dashed lines
  represent the spectra before C2 and C5 subtraction.  In each panel
  the thick smooth line represents the laboratory ice mixture
  HCOOH:H$_2$O:CH$_3$OH=12:100:40 at $T=15$ K after removal of the
  H$_2$O bending mode \citep{bis07}.  The laboratory spectra are
  scaled such that the integrated optical depth of the 7.25 \mum\ band
  matches the observations. Note that the laboratory spectrum of the
  7.25 \mum\ band is wider than shown in \citet{bis07}. This is due to
  a different baseline choise at the steep edge of the strong C-H
  deformation mode of CH$_3$OH in the laboratory
  data.}~\label{f:hcooh}
\end{figure}

The main vibrational modes of formic acid (HCOOH) occur in the 3-4
\mum\ region (O-H and C-H stretch), at 5.85 \mum\ (C=O stretch), at
7.25 \mum\ (C-H deformation), and at 8.1 \mum\ (C-O stretch).
Previously, the 7.25 \mum\ band was detected toward massive YSOs,
while the 5.85 \mum\ band, in addition to H$_2$O, could explain most
of the 6 \mum\ absorption \citep{sch99}.  The 7.25 \mum\ band offers
the most secure identification, even though its intrinsic strength
($A=2.6\times 10^{-18}$ cm/molecule) is a factor of $\sim$25 weaker
than the 5.85 \mum\ band.  It is least blended with other features and
it is unambiguously detected ($\geq\ 3\sigma$) toward 9 low mass YSOs
(Fig.~\ref{f:7_8um}; Table~\ref{t:colden}).  With the aid of
laboratory analogs \citep{bis07}, the observed strength of the 7.25
\mum\ feature is used to determine the contribution of HCOOH to the
6.0 \mum\ band. Pure HCOOH, and mixtures with H$_2$O show 7.25 \mum\
features that are at least a factor of 2 wider than the observed band
(FWHM$\sim$15 \waven), and that peak at too short wavelengths.
\citet{sch99} and recently \citet{bis07} note that the band narrows
and shifts to longer wavelengths in tertiary mixtures, specifically in
H$_2$O:CH$_3$OH:HCOOH=100:40:12 (Fig.~\ref{f:hcooh}).  Such a mixture
would be consistent with the detections and upper limits of solid
CH$_3$OH (typically $N$(CH$_3$OH)$>N$(HCOOH) toward YSOs;
Table~\ref{t:colden}).  For the purpose of this analysis the
laboratory and interstellar data are matched by scaling their
integrated 7.25 \mum\ optical depths.  For these sources, the HCOOH
column densities are 1-5\% with respect to solid H$_2$O
(Fig.~\ref{f:abun}) and 50-100\% of component C1 can be explained by
the C=O stretching mode of HCOOH (Fig.~\ref{f:hcooh}).  This wide
range encompasses the large uncertainty in the band strength: the
laboratory 5.85 \mum\ band is stronger by 60\% in H$_2$O mixtures
compared to pure HCOOH.  For many sources in which the 7.25 \mum\ band
is not detected ($<3 \sigma$), a stronger constraint on the HCOOH
column density is provided by the strength of the C1 feature.
Relative HCOOH abundances of a few \% are derived by using the C=O
band strength of the H$_2$O-rich mixtures.  Finally, in most cases
neither the series of shallow, broad HCOOH bands in the 3-4 \mum\ O-H
and C-H stretching mode region, nor a band at 8.1 \mum, blended with
the edge of the prominent 9.7 \mum\ silicate feature, provide stronger
constraints to the HCOOH column density.

\subsubsection{H$_2$CO}~\label{sec:h2co}

As was noted for massive YSOs (e.g. \citealt{kea01b}), the C1
component shows a sharper edge on the short wavelength side than can
be explained by HCOOH alone.  The C=O stretching mode of H$_2$CO is a
viable candidate, and, assuming an integrated band strength of
$A=9.6\times 10^{-18}$ cm/molecule \citep{sch93}, substantially better
fits to component C1 are obtained by adding typically an abundance of
6\% of H$_2$CO with respect to H$_2$O (Fig.~\ref{f:labcombo}).
Accurate H$_2$CO abundances and uncertainty estimates are hard to
establish due to the strong blend with HCOOH.  This abundance estimate
is consistent with the non-detection of the C-H stretching bands at
3.34, 3.47, and 3.54 \mum\ and leads to a contribution of 10-35\% from
H$_2$CO to component C1.  At the quoted typical abundance, the C-H
bending mode vibration of H$_2$CO at 6.68 \mum\ contributes $\leq$15\%
to component C3 (assuming $A=3.9\times 10^{-18}$ cm/molecule from
\citealt{sch93}).

\subsubsection{NH$_3$}~\label{sec:nh3}

The abundance of solid NH$_3$ has long been a matter of debate,
because all of its absorption bands overlap with much stronger bands
of H$_2$O and silicates. A combined in-depth analysis of all these
features is therefore required and accurate column densities will be
presented elsewhere (S. Bottinelli et al., in prep.).  In this work,
NH$_3$ column densities accurate to within a factor of a few are
sufficient. The strength of the 9.0 \mum\ umbrella mode is used to
calculate the strength of the 6.16 \mum\ N-H bending mode and its
contribution to component C2. The inflection seen at 9.0 \mum\ in
several sources is fitted with a Gaussian on optical depth scale and a
detection of NH$_3$ is claimed if the measured peak position and FWHM
width are within 8.91-9.05 \mum\ and 0.20-0.45 \mum, respectively.
Thus, NH$_3$ is positively identified in 17 low mass YSOs.  Column
densities were calculated using an integrated band strength of
$1.3\times 10^{-17}$ cm/molecule for NH$_3$ mixed in an H$_2$O ice
\citep{ker99}, leading to values of typically 3-8\% with respect to
H$_2$O. At these abundances, 10-50\% of the C2 component is due to
solid NH$_3$.

\subsubsection{HCOO$^-$}~\label{sec:hcoom}

The formate ion (HCOO$^-$) is an interesting candidate to consider
because its C=O stretching mode at 6.33 \mum\ overlaps with the C2
component, its C-H deformation mode coincides with the 7.40 \mum\ band
detected in several sight-lines (Fig.~\ref{f:7_8um}), and because it
is chemically related to the HCOOH species discussed above
\citep{sch99}.  Its identification is tentative, however, since
acetaldehyde (CH$_3$HCO) has a feature at 7.40 \mum\ as well, but not
near the C2 component. If HCOO$^-$ were the carrier of the 7.40 \mum\
band, its column density would be $\sim$0.3\% with respect to solid
H$_2$O, an order of magnitude less than HCOOH, and its contribution to
component C2 would be $\leq 20$\%.

\subsubsection{Composite Laboratory Spectrum}

\begin{figure}
\includegraphics[angle=90, scale=0.50]{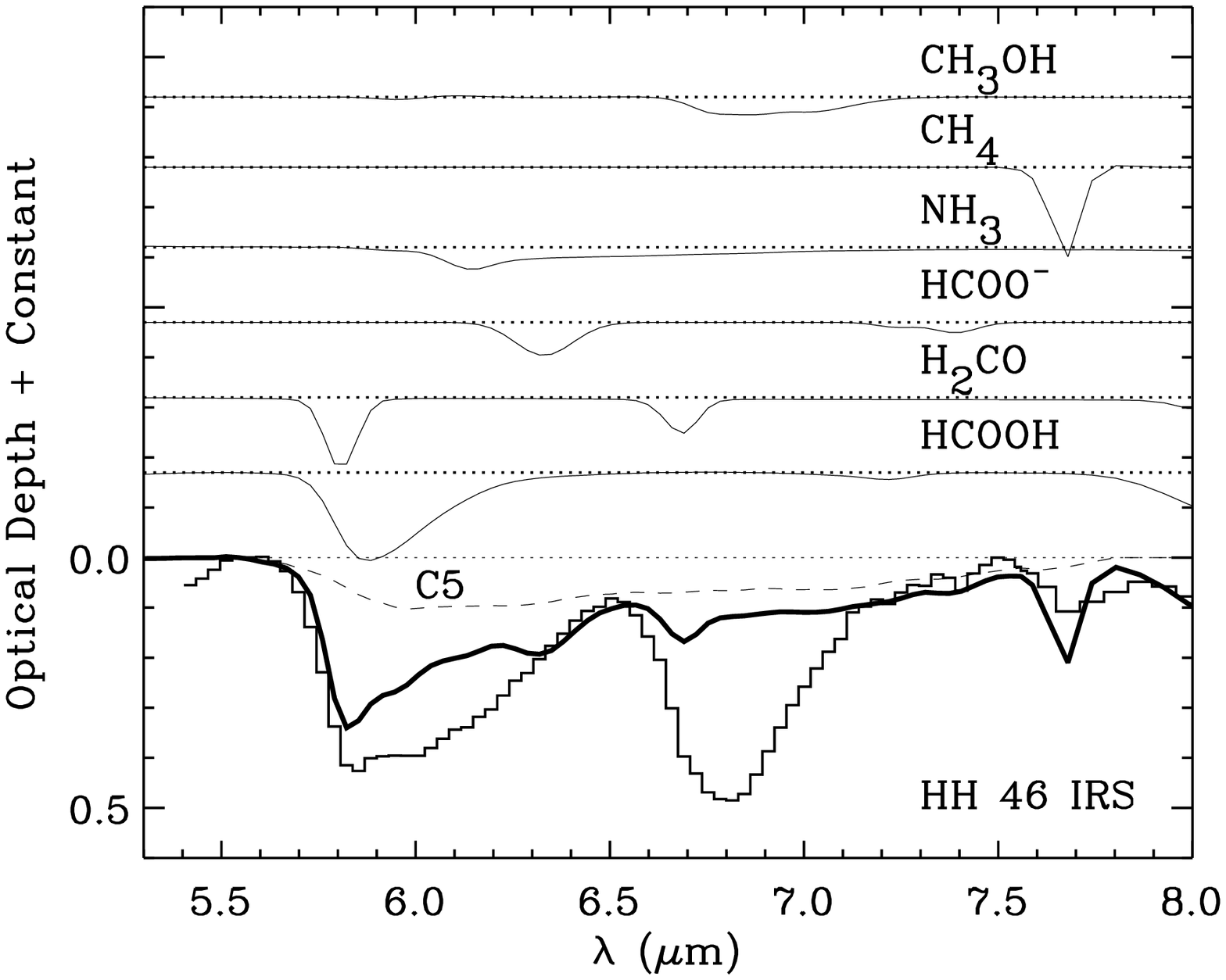}
\caption{Simple species contributing to the 6 and 6.85 \mum\ bands of
  the low mass YSO HH~46 IRS, after H$_2$O subtraction
  ($N$(H$_2$O)=82.8$\times 10^{17}$ \sqcm). Laboratory spectra of
  (from top to bottom) solid CH$_3$OH, CH$_4$, NH$_3$, HCOO$^-$,
  H$_2$CO, and HCOOH were scaled to column densities derived
  independently elsewhere in the spectrum: 5.6\% (CH$_3$OH), 2.2\%
  (CH$_4$), 12.7\% (NH$_3$), 0.4\% (HCOO$^-$), 6.0\% (H$_2$CO), and
  2.0\% (HCOOH) with respect to H$_2$O.  These spectra were
  subsequently added to the underlying, unidentified C5 absorption
  component (dashed line) and the result is shown as a solid thick
  line. Clearly, the prominent 6.85 \mum\ band is not explained by
  these particular species.}~\label{f:labcombo}
\end{figure}

To illustrate that the independently derived column densities of solid
HCOOH, HCOO$^-$, H$_2$CO, NH$_3$, and CH$_3$OH are indeed consistent
with the 5-7 \mum\ observations, laboratory spectra of these ices are
scaled and subsequently added (Fig.~\ref{f:labcombo}).  While no
attempt is made to match the band profiles in detail (profiles are
highly dependent on interactions in the ice matrix and a detailed
analysis awaits further laboratory work), the agreement is
encouraging. A considerable fraction of the interstellar 5-8 \mum\
absorption consists of a complex blend of features from these simple
ices.  In particular, they explain most of the observed strength of
component C1 (50-100\%), much of C2 (30-70\%) and some of C3
($\leq$30\%).  At the same time, Fig.~\ref{f:labcombo} also shows that
component C5 and most of the prominent 6.85 \mum\ absorption band (C3
and C4) are not accounted for.  NH$_4^+$ is a candidate for the latter
(\S\ref{sec:disc3c4}), but its column density cannot be determined
independently and thus this species is not included in this plot.

\subsection{Correlations}~\label{sec:corr}

\begin{figure}
\includegraphics[angle=90, scale=0.50]{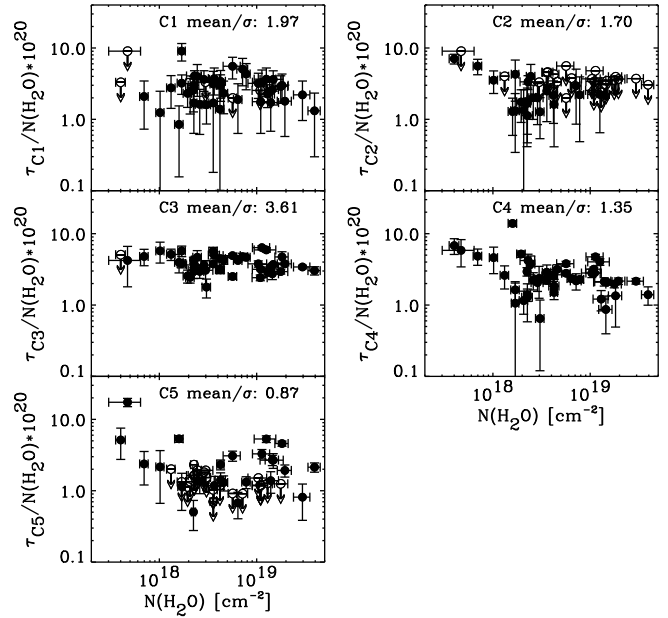}
\caption{Peak optical depths of the 5--8 \mum\ components (Fig.
  \ref{f:decomp}), normalized to $N$(H$_2$O), plotted as a function of
  $N$(H$_2$O).  Error bars are 1$\sigma$, but open symbols with arrows
  attached refer to 3$\sigma$ upper limits. At the top right of each
  panel the ratio of the mean over the standard deviation of the
  sample (excluding upper limits) is indicated.  Component C3,
  i.e. the short-wavelength component of the 6.85 \mum\ feature,
  correlates significantly better with $N({\rm H_2O})$ than the other
  components do.  Of the components that produce the 6.0 \mum\ excess,
  C1 correlates better with $\rm H_2O$ than does C5. Note the
  increased strengths of C2 and C4 at low $N({\rm
  H_2O})$.}~\label{f:compcorrel}
\end{figure}

To obtain further insight into the origin of the C1-C5 components,
their strengths were normalized to the H$_2$O column density for the
entire sample and plotted in Fig.~\ref{f:compcorrel}.  As a measure of
the degree of correlation the ratio of the mean values and standard
deviations were calculated (excluding upper limits), yielding values
of 1.97, 1.70, 3.61, 1.35, and 0.87 for C1-C5 respectively. Thus, of
all components, C3 correlates best with H$_2$O, followed by C1. The C2
and C4 components show more scatter, and, most interestingly, their
strengths are largest at the lowest H$_2$O abundances.  Component C5
correlates poorest with H$_2$O, especially when taking into account
the tight 3$\sigma$ upper limits at the bottom end of the
distribution.  While, of all components, C5 is most susceptible to
errors in the continuum determination, the C5/$N$(H$_2$O) variation is
too large to be explained by that alone (\S\ref{sec:60}).  Therefore,
the large variations in the 6.0 \mum\ excess ($E_6$;
Fig.~\ref{f:h2oh2o}) can be ascribed primarily to variations in
component C5, and much less to C1.

The different degrees of correlation of the C1-C5 components with
H$_2$O indicate that they have different origins.  To further
elucidate their nature, their strengths were correlated with a number
of tracers of the chemical or physical processing history of the ices.
A well-known tracer is the ratio of broad and narrow 4.7 \mum\ CO ice
components (e.g. \citealt{chi98}), often referred to as `polar' and
`apolar' ices due to the mixing of CO with high and low dipole moment
species, respectively.  The apolar ices have a much lower sublimation
temperature than the polar ices ($T_{sub}=20$ versus 90 K) and thus
their abundance ratio traces thermal processing.  Higher temperatures
are traced by the 15 \mum\ CO$_2$ bending mode. Characteristic double
peaks are visible in the crystalline ices produced after heating
CO$_2$:H$_2$O mixtures to $T\geq 50$ K \citep{ger99}, and the fraction
of crystalline CO$_2$ ice so derived for many sources in our sample is
presented in \citet{pon08}.  Actual dust temperatures can be derived
from far-infrared colors, but to trace the coldest dust components
($T<$20 K) photometry at wavelengths $>100$ \mum\ is required.  For
massive YSOs this was provided by ISO/LWS spectra, and the
far-infrared color was found to correlate with indicators of thermal
processing \citep{boo00_1}.  Unfortunately, such photometry is not
available for most of our sample of low mass YSOs.  Tracers of
ultraviolet radiation and cosmic ray fluxes are even harder to obtain.
Traditionally, the strength of the 4.62 \mum\ absorption band has been
used for this, since the carrier, an 'XCN'-bearing species, was
thought to be produced by the effects of UV photons or cosmic rays
(e.g.  \citealt{pen99}). Recent work shows, however, that the
potential carrier OCN$^-$ can be produced by heating an
HNCO-containing ice formed at cold conditions by grain surface
chemistry \citep{bro04}.  Thus, in this case the strength of the 4.62
\mum\ band is in fact another tracer of thermal evolution.

\begin{figure}
\includegraphics[angle=90, scale=0.60]{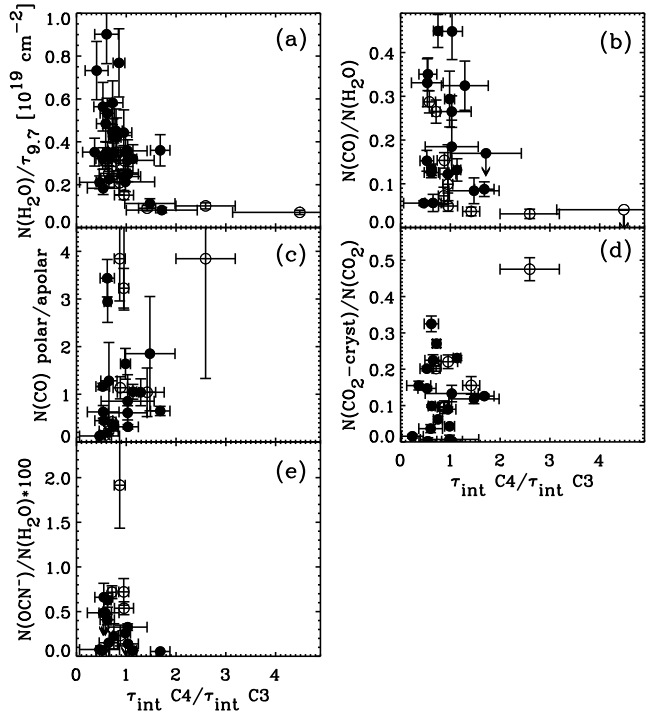}
\caption{The ratio of the integrated strengths of components C4 and C3
  (red and blue components of the 6.85 \mum\ band) plotted against
  possible ice or dust temperature tracers. Along the $y-$axis are
  plotted: ({\bf panel a}) a measure of the solid H$_2$O abundance,
  ({\bf panel b}) a measure of the solid CO abundance, ({\bf panel c})
  the polar/apolar solid CO ratio, ({\bf panel d}) the fraction of
  crystalline CO$_2$ measured from the 15 \mum\ bending mode, and
  ({\bf panel e}) the relative abundance of the carrier of the 4.62
  \mum\ feature, OCN$^-$. Low mass YSOs are indicated by filled dots,
  massive YSOs by circles, and background stars by star symbols.
  Panels (a) and (b) show that sources with prominent C4 components
  have low ice (H$_2$O and CO) abundances.}  ~\label{f:corr685}
\end{figure}

The prominent 6.85 \mum\ absorption band consists of independent short
and long-wavelength components (C3 and C4; Fig.~\ref{f:decomp}), whose
integrated optical depth ratio varies considerably (C4/C3=0.5-4). This
essentially reflects abundance variations of the carrier of the C4
component, because C4 correlates more poorly with the H$_2$O column
density than does the C3 component (Fig.~\ref{f:compcorrel}). The
C4/C3 ratio is plotted against the observables discussed above, and
while all of the observables show a large spread at C4/C3$\le$1.2,
sources with stronger C4 components all have low solid H$_2$O and CO
abundances (panels (a) and (b) of Fig.~\ref{f:corr685}).  These
include the low mass YSOs IRAS 03301+3111 and DG Tau B, and the
massive YSOs Mon R2 IRS3 and S140 IRS1.  The weakness of the ice bands
in these particular sources hampers the correlation study with some of
the above-mentioned observables: the ratio of the polar/apolar solid
CO components and the fraction of crystalline CO$_2$ cannot be
determined if these bands are not or weakly detected.  Similarly, only
a handful of OCN$^-$ abundance determinations are available for these
sources.  It is thus concluded that the C4/C3 ratio is enhanced in
environments that have low ice abundances. This is in agreement with
earlier work on massive YSOs \citep{kea01b}, which showed an increase
of the ratio in lines of sight with warmer dust temperatures.
Potential carriers of the C4 and C3 components are discussed in
\S\ref{sec:disc3c4}.

\begin{figure}
\includegraphics[angle=90, scale=0.60]{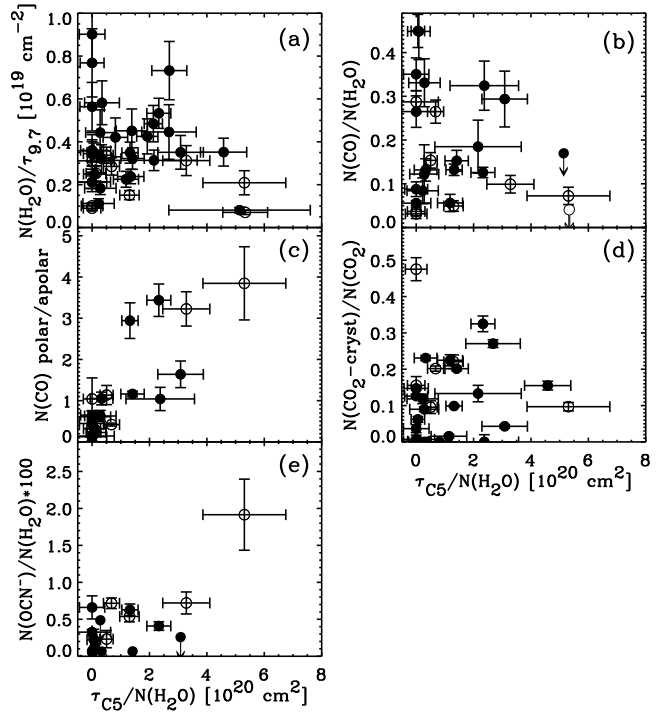}
\caption{The peak strength of component C5 scaled to $N$(H$_2$O)
  plotted against the same ice or dust temperature tracers as in
  Fig.~\ref{f:corr685}. Possibly, enhanced polar CO (c) fractions
  are observed at larger C5 strengths.}~\label{f:corrc5}
\end{figure}

The strength of the enigmatic C5 component varies widely for both
low and high mass YSOs.  Toward massive YSOs, a trend of increasing
strength of the C5 component with the OCN$^-$ abundance (4.62 \mum\
feature) was claimed \citep{gib02}, but in the current sample this
trend is rather weak and is limited by the small number of sources
with OCN$^-$ detections (panel (e) in Fig.~\ref{f:corrc5}).  A
stronger trend is observed with the ratio of the polar and apolar CO
components: sources with strong C5 components show larger
polar/apolar CO ratios (panel (c) in Fig.~\ref{f:corrc5}).  No
correlations are observed with the H$_2$O, CO, and crystalline CO$_2$
abundances. Potential carriers are further discussed in
\S\ref{sec:disc5}

\section{Discussion}~\label{sec:dis}

Long-standing key questions in astrochemistry concern the composition
of the interstellar ices and the degree to which the composition of
the ices is altered by heating and by the impact of energetic photons
and cosmic rays.  The 3-30 \mum\ spectra presented in this paper
extend studies addressing these questions from the previously well
analyzed massive YSOs to low mass YSOs, as well as to background stars
tracing quiescent cloud material.  The 5--8 \mum\ region exhibits a
very rich absorption spectrum, which, apart from the O-H bending mode
of H$_2$O, consists of at least 5 independent components. While most
of component C1, much of C2 and part of component C3 is due
to the simple species CH$_3$OH, HCOOH, NH$_3$, and H$_2$CO, the origin
of a substantial amount of absorption is still unclear, and this is
discussed in more detail below.

\subsection{Constraints on the Carrier(s) of Components C3 and
  C4 (the 6.85 \mum\ Band)}~\label{sec:disc3c4}

The 6.85 \mum\ band was decomposed into `blue' and `red' components
(C3 and C4). While C3 correlates tightly with the H$_2$O abundance,
the relative strength of the C4 component increases at the lowest
H$_2$O abundances (Fig.~\ref{f:compcorrel}).  This could be explained
by the formation of a new species as the ices are heated and sublimate
or it could be a pre-existing species that is less volatile than
H$_2$O whose absorption band (C4) prevails at lower H$_2$O abundances.
In the first scenario heating must play a role, and in the second it
may.  The detection of the C4 component in all YSOs and toward
background stars favors the second scenario, in which the carrier is
not formed by heating for its carrier exists already in the earliest
stages of chemical evolution, i.e. in quiescent clouds.  Higher dust
temperatures in the vicinity of YSOs lead to sublimation and lower
H$_2$O and CO abundances (Fig.~\ref{f:corr685}) and the abundance of
the carrier of the C3 component is reduced more rapidly than that of
the C4 component.

Does the carrier of the C4 component survive in the diffuse
interstellar medium? A band at 6.85 \mum\ is detected in lines of
sight toward the Galactic Center (GC), tracing a large column of
diffuse dust \citep{chi00}. However, its depth relative to the 9.7
\mum\ silicate band ($\tau_{6.85}/\tau_{9.7}\sim 0.014$) is a factor
of 5 less compared to the C4 component in the YSOs (factor of 10
compared to C3).  The weak 6.85 \mum\ band observed toward the GC
sources is attributed to the C-H asymmetric deformation mode of
aliphatic hydrocarbons \citep{chi00}.  Such aliphatic hydrocarbons
have strong C-H stretching modes with distinct peaks at 3.38, 3.42,
and 3.48 \mum. A band at 3.47 \mum\ is commonly detected toward YSOs
\citep{bro99}, but it does not show these distinct peaks and its depth
relative to the 6.85 \mum\ band is a factor of 3 weaker compared to
the GC sources. Thus, although the carrier of the C4 component
observed in our sample is (probably) less volatile than ices, it does
not survive in the diffuse ISM. Conversely, the hydrocarbons observed
in the diffuse ISM are not the carrier of the C4 component toward
YSOs.

In a scenario proposed by \citet{sch03} both the C3 and C4
components arise from an NH$_4^+$-containing ice. Its abundance would
be typically 7\% with respect to H$_2$O (Table~\ref{t:colden} and
Fig.~\ref{f:abun}), assuming the integrated band strength of
$4.4\times 10^{-17}$ cm/molecule measured in \citet{sch03}.  NH$_4^+$
ions can be produced either by acid-base reactions at low temperature
(e.g.  from NH$_3$:HNCO at 10 K; \citealt{rau04b}) or by ultraviolet
or cosmic ray irradiation of simple mixtures (e.g.
H$_2$O:CO$_2$:NH$_3$:O$_2$).  The NH$_4^+$ band that is produced
shifts to longer wavelengths upon heating, i.e. from the peak
wavelength of the C3 component to the peak wavelength of the C4
component.  After additional energetic processing and H$_2$O
sublimation a salt is formed. Salts do not sublimate until
temperatures of $\sim$200 K (e.g. \citealt{sch03}), significantly
higher than the H$_2$O sublimation temperature of 90 K in space.  The
processed laboratory spectra show absorption at 6.0 \mum\ due to
anions (NO$_2^-$, NO$_3^-$, and HCO$_3^-$), some of which overlaps
with the C2 components centered at $\sim$6.15 \mum\ that are
observed in the YSOs with strong C4 components (Fig.~\ref{f:resid};
\S\ref{sec:disc2}).  While these properties are consistent with the
observations of YSOs, the presence of the C4 component toward all
YSOs and background stars is hard to explain.  In the experiments of
\citet{sch03} an NH$_4^+$ temperature of almost 200 K is required to
match the peak position of this component. Such a high temperature can
be excluded for many sources in the sample.  Perhaps the impact of
cosmic rays is responsible for the shift in peak position.  Subsequent
heating by nearby protostars leads to the observed increase in the
C4/C3 ratio and eventually to the formation of a salt.


\subsection{Potential Carrier(s) of Component C5}~\label{sec:disc5}

The enigmatic C5 component that stretches from 5.8 to $\sim$8.0 \mum\
is not due to NH$_3$, H$_2$CO, HCOOH, HCOO$^-$, or CH$_3$OH, because
these species do contribute to components C1-C4 (\S\ref{sec:carr}).
Also, no correlation is found between the strengths of the C5 and the
6.85 \mum\ red (C4) components.  Therefore, these have unrelated
origins.  Heating might be involved in the formation of the carrier(s)
of the C5 component (Fig.~\ref{f:corrc5}; \S\ref{sec:corr}).
Previously, three different origins were considered in the literature.

First, it should be noted that the shape of the average C5 component
resembles that of the spectrum of solid H$_2$O (Fig.~\ref{f:compex}).
It is particularly reminiscent of heated H$_2$O ($T>120$ K), because
of the weakness of the 6.0 \mum\ peak with respect to the long
wavelength wing (e.g.  \citealt{mal98}). While the shape of the 3
\mum\ band indicates that heated ices are present along several lines
of sight (e.g., RNO 91), its depth is, by the very definition of the
C5 component, insufficient.  Possibly, due to the effects of light
scattering at the shorter wavelengths, a more direct path to the warm
inner regions of the YSO is observed at 6 \mum\ compared to 3 \mum.
Indeed, Monte Carlo radiative transfer models of a circumstellar disk
show that at inclination angles of $70-74^o$ the depth of the 6.0
\mum\ H$_2$O band is enhanced by a factor of 10 with respect to that
at 3.0 \mum\ \citep{pon05}. At higher inclinations, the enhancement is
a factor of 3--4, while at lower inclinations the effect is
negligible.  Thus, the C5 component may be attributed to heated H$_2$O
ice for some sources in our sample; for example, RNO 91 is known to
possess an inclined disk \citep{wei94} and has one of the most
prominent C5 components.  At the other extreme, the weakest C5
components are observed toward background stars (Table~\ref{t:tau}).
Indeed, in these lines of sight the effect of scattering on the
relative H$_2$O band depths is expected to be small due to the
uniformity of the foreground clouds.  For many YSOs with C5
absorption, however, independent observational evidence indicates that
the scattering scenario is unlikely.  For example, the CH$_3$OH column
density derived from the 3.53 \mum\ C--H stretching mode agrees well
with that derived from the 9.7 \mum\ C--H bending mode
(\S\ref{sec:ch3oh}; \citealt{sch91}).  Also, for high mass YSOs the
4.25 and 15 \mum\ bands provided solid CO$_2$ column densities that
were in excellent agreement with each other \citep{ger99}, and in one
line of sight also with the 2.7 \mum\ CO$_2$ combination mode
\citep{kea01a}.  Thus, for these YSOs other explanations are required.

A second explanation of the C5 component could be the blend of
negative ions suggested in \citet{sch03}. Ammonium (NH$_4^+$) and the
anions HCO$_3^-$, NO$_3^-$, and NO$_2^-$ are produced after UV
photolysis or cosmic ray bombardment of H$_2$O/CO$_2$/NH$_3$/O$_2$ ice
mixtures. NH$_4^+$ has a strong band at 6.85 \mum\ (i.e. components
C3 and C4 in this work), but the bands of the anions overlap
such that only one very broad feature is observed in the 5-8 \mum\
range.  While one may expect that in this scenario the C3/C4 and
C5 components correlate well with each other, the observations show
they do not. Notably, the background stars have strong 6.85 \mum\
components and little C5 absorption. The NH$_4^+$ ion can be
produced after warming up a simple ice mixture too, however, while the
HCO$_3^-$, NO$_3^-$, NO$_2^-$ anions may need more energetic input
\citep{sch03}.  This may explain the large variation in the strength
of the C5 component.

Third, extensive processing of simple ices by UV light or cosmic rays
produces an organic refractory residue \citep{gre95}, whose blended
O-H and C-H stretching and bending modes produce a feature similar to
the C5 component \citep{gib02}. Both this scenario and the blend of
ions discussed above require a correlation of the strength of the C5
component with independent indicators of the presence of strong UV or
cosmic ray fluxes, which are currently unavailable.  Also, both the
organic residue and the salts are less volatile than the ices---they
sublimate at temperatures $\leq$500 K in the laboratory
(\citealt{sch03}, \citealt{gre85}; $T_{sub}$=180 K for H$_2$O ice).
Therefore, one may expect C5 to be the main absorption component in
the 5--8 \mum\ region toward some YSOs where all H$_2$O ices have
evaporated.  Although in several sources the C5 component is the
dominant contributor, it always is observed together with H$_2$O
(except possibly toward L1448 IRS1).  Some H$_2$O may be trapped in
the salt or organic residue at high temperatures, and some may be
present in unprocessed foreground material.  It is also important to
note that, similar to C4 (\S\ref{sec:disc3c4}), the C5 component is
not observed in the diffuse ISM, confirming that their carrier(s) are
not as refractory as silicates ($T_{sub}\sim$1500 K).

Finally it is noted that the C5 component may have a different origin
in different lines of sight. An in-depth analysis of the physical
conditions in specific lines of sight is required to discriminate
between the effects of disk orientation and the various ice processing
mechanisms.

\subsection{Potential Carrier(s) of Component C2}~\label{sec:disc2}

The C2 component spans the 6.0-6.4 \mum\ wavelength range and is
likely due to at least three different carriers. The N-H bending mode
of NH$_3$ at 6.15 \mum\ contributes 10-50\% to this band
(\S\ref{sec:nh3}). Another likely contributor is H$_2$O.  Although the
spectrum of pure, amorphous H$_2$O was subtracted from the 5-8 \mum\
region before determining the various component strengths and
profiles, the spectrum of interstellar H$_2$O may deviate from that of
a pure, amorphous H$_2$O ice.  The O-H bending mode of H$_2$O shifts
to longer wavelengths, becomes significantly narrower, and its peak
strength increases if the hydrogen bonding network in bulk amorphous
H$_2$O ice is broken by the presence of other species \citep{ehr96,
  obe07}.  CO$_2$ is a realistic species, because it is abundant
(15-40\% with respect to H$_2$O) and a significant fraction is known
to be mixed with H$_2$O \citep{pon08, ger99}. At sufficiently high
CO$_2$ concentrations, the H$_2$O is located in small clusters,
leading to monomer, dimer, and multimer bondings that produce a strong
peak at 6.12 \mum. The optical depth of the peak has been calculated
for the YSO B5 IRS1 \citep{obe07} and the background star Elias 16
\citep{kne05} assuming the H$_2$O ice along these lines of sight
consists of two components: pure H$_2$O and the mixture
H$_2$O:CO$_2$=2:1.  It was found that when 25\% of the H$_2$O is
present in the latter, the 3 and 6 \mum\ bands of H$_2$O and the 15
\mum\ band of CO$_2$ are fitted simultaneously.  Comparison with the
C2 component in these sources shows that 20-50\% could be due to this
effect.  Thus the short wavelength side of the C2 component is both
due to H$_2$O:CO$_2$ mixtures and due to NH$_3$.  For the
long-wavelength side, the C=O stretching mode of the HCOO$^-$ anion
was previously considered, contributing $\leq$20\% to the C2 strength
(\S\ref{sec:hcoom}).  Other anions, such as NO$_2^-$, NO$_3^-$, and
HCO$_3^-$ (\S\ref{sec:disc3c4}), may contribute as well.  These
particular anions may be produced in the energetic processing of
simple ices \citep{sch03}, along with NH$_4^+$.  Indeed, one of the
most interesting observational aspects of the C2 component is that it
becomes more pronounced for sources with stronger C4 components and
lower H$_2$O abundances (Figs.~\ref{f:resid} and~\ref{f:compcorrel}).
Overall the two components do not correlate very well, however, which
is consistent with the result that three carriers of different nature
contribute to the C2 component.

\subsection{Astrochemical Implications}~\label{sec:dischem}

CH$_3$OH shows larger abundance variations with respect to H$_2$O than
other solid state species discussed in this work (HCOOH, H$_2$CO,
NH$_3$). Tight 3$\sigma$ upper limits of a few \% are reported for a
number of YSOs, but detections above the 10\% level for others
(Table~\ref{t:colden} and Fig.~\ref{f:abun}). The solid HCOOH
abundance is typically a factor of 2-3 lower than CH$_3$OH and does
not show such large variations. If both CH$_3$OH and HCOOH are formed
from CO by grain surface reactions, the lower HCOOH abundances are
understood by the low atomic O/H ratio in the accreted gas, and by the
competition for O to react to CO$_2$ rather than with the formyl
radical (HCO) to form HCOOH \citep{tie82}.  Indeed, large solid CO$_2$
abundances (15-40\% with respect to H$_2$O) are observed for a wide
variety of YSOs \citep{pon08}. Grain surface chemistry does indeed
appear a more likely formation mechanism than UV processing of the
ices.  First, while processing of simple ices by UV photons leads to
CH$_3$OH formation, this process is not efficient enough to explain
the enhanced abundances toward massive YSOs \citep{dar99}, while for
low mass YSOs the UV flux is likely too low regardless.  Second, HCOOH
and CH$_3$OH are detected even in the most obscured sight-lines in the
sample. In fact, enhanced CH$_3$OH abundances are detected exclusively
in some of the most embedded YSOs: the low mass YSOs SVS 4-5, B1-b,
and IRAS 15398-3359 and the massive YSOs W33A and GL7009S
\citep{dar99}.  SVS 4-5 is known to trace the envelope of the
foreground Class 0 object SMM4 \citep{pon04}. On the other hand,
toward a few background stars tracing quiescent cloud material a low
upper limit to the CH$_3$OH abundance was determined
(Table~\ref{t:colden}; \citealt{chi96, kne05}).  Thus it is suggested
that, in the grain surface chemistry scenario, the CH$_3$OH abundance
is enhanced at a time during the collapse (of some clouds) when the
atomic O needed to form H$_2$O and CO$_2$ is exhausted, while still
sufficient CO and H are available to form CH$_3$OH \citep{obe08}.

While grain surface chemistry likely forms the simple neutral species,
different mechanisms are required to produce the ions possibly
responsible for several of the components in the 5-8 \mum\ range.  If
the observed C3 and C4 components are indeed due to NH$_4^+$, its
abundance is typically 7\% with respect to H$_2$O (Fig.~\ref{f:abun}).
The C4 component is observed even in the coldest sight-lines.  Indeed,
low temperature acid-base chemistry can efficiently produce NH$_4^+$
\citep{rau04b}.  The abundance of negative counter ions needs to be
addressed.  While the C5 component could hide the necessary negative
ions \citep{sch03}, its strength does not correlate with the C3/C4
components.  Further investigations on the presence of ions in the
ices are crucial to understand the ice chemistry. Their physical
properties are different from neutral species.  Ions have higher
sublimation temperatures, and thus the salts that are formed after the
volatile species (e.g. H$_2$O and CO$_2$) have sublimated are able to
survive at lower extinctions, and then are subjected to harsher
conditions, forming more complex species. For example, the
biologically interesting species urea (H$_2$NCONH$_2$) is formed after
the UV irradiation of a NH$_4^+$OCN$^-$ salt \citep{rau04a}.

\section{Conclusions and Future Work}~\label{sec:concl}

The present work extends the study of ices over the full 3-20 \mum\
wavelength range from the previously well studied massive YSOs
($\sim10^5$L$_{\odot}$) to low mass YSOs ($\sim$1L$_{\odot}$). The
following new insights were obtained:

\begin{itemize}

\item Absorption features are detected at 6.0 and 6.85 \mum\ in all
  sources, including high- and low-mass YSOs and background stars
  tracing quiescent cloud material. An empirical decomposition shows
  that the 5-7 \mum\ absorption complex consists, in addition to the
  H$_2$O bending mode, of a combination of at least 5 independent
  absorption components.

\item In a subset of sources additional weak features are detected at
  7.25, 7.40, and at 9.0 and 9.7 \mum\ on top of the prominent Si-O
  stretching band of silicates. These features are associated with
  CH$_3$OH, HCOOH, NH$_3$, and possibly HCOO$^-$ and indicate
  abundances of 1-30\%, 1-5\%, 3-8\%, and 0.3\% with respect to
  H$_2$O, respectively.  The large source-to-source solid CH$_3$OH
  abundance variations are likely a result of the conditions at the
  time of grain surface chemistry.

\item Component C1 (5.7-6.0 \mum) is mostly explained by solid HCOOH
  and H$_2$CO at abundances with respect to solid H$_2$O of typically
  1-5\% and $\sim$6\%, respectively.

\item Component C2 (6.0-6.4 \mum) also likely arises from a blend of
  several species.  Solid NH$_3$ can account for 10-50\% of the
  absorption. A similar amount can be attributed to absorption by
  monomers, dimers, and small multimers of H$_2$O mixed with a
  substantial amount of CO$_2$. The long-wavelength side of C2 is
  potentially due to anions produced by acid-base chemistry (e.g.
  HCOO$^-$) or energetic processing.  Indeed, a weak correlation with
  C4 suggests the carriers of these bands (possibly salts) are
  related.

\item Components C3 and C4, the 6.85 \mum\ band, show the same
  characteristics for low mass YSOs and background stars as was
  previously found for massive YSOs by \citet{kea01b}.  Their ratio is
  empirically found to be dependent on dust temperature. The carrier
  of C4 is likely less volatile than that of C3 and than H$_2$O, but
  it is not observed in the diffuse ISM.  These characteristics are
  also consistent with both components being due to NH$_4^+$ (ammonium
  ion). The detection of strong 6.85 \mum\ bands toward deeply
  embedded YSOs and background stars requires a production at low
  temperature (acid-base chemistry or processing by cosmic rays),
  given the lack of heating sources and stellar UV fields.

\item The origin of the very broad component C5 (5-8 \mum) is least
  understood. It is quite strong in several low and high mass YSOs,
  absent in others and so far undetected toward background stars. It
  is possibly related to thermal processing, as a weak correlation
  with the ratio of the polar/apolar CO components is observed and its
  shape resembles that of warm H$_2$O.  A blend of absorption by ions,
  as proposed by \citep{sch03}, which would require both heating and
  energetic processing of the ices, cannot be excluded. The latter
  could also lead to an organic residue, whose shape resembles that of
  the C5 component as well.

\item Weak correlations are found between the absorption components in
  the 5-8 \mum\ range and tracers of physical conditions.  A more
  thorough understanding of the conditions, the source geometry and
  local radiation fields in specific lines of sight is required to
  further constrain the nature of the carriers (in particular for
  components C2-C5) and the importance of thermal and energetic
  processing.  To what degree does scattering play a role? What are
  the dust temperatures, cosmic ray fluxes, UV radiation fields and
  time scales as a function of distance along the line of sight?
  Measurements of source fluxes above wavelengths of 100 \mum\ with
  the Herschel Space Observatory will be valuable, as well as
  ground-based measurements of the 4.62 \mum\ band of OCN$^-$ in more
  lines of sight.  In addition, further laboratory work, in particular
  on the effect of cosmic rays on the ices, especially for the
  identification of components C3 and C4, is required. Finally, study
  of the 5-8 \mum\ features toward a larger sample of background stars
  is crucial to make further progress in the identification and
  processing history of interstellar ices.

\end{itemize}

\acknowledgments We thank the anonymous referee for thoughtful
comments on the manuscript and Alberto Noriega-Crespo for help in
identifying the emission lines.  Support for this work, part of the
Spitzer Legacy Science Program, was provided by NASA through contracts
1224608, 1230779, 1230782, 1256316, and 1279952 issued by the Jet
Propulsion Laboratory, California Institute of Technology, under NASA
contract 1407.  Astrochemistry in Leiden is supported by a NWO Spinoza
grant, a NOVA grant, and by the European Research Training Network
``The Origin of Planetary Systems" (PLANETS, contract number
HPRN-CT-2002-00308).  The work of JKJ is supported by NASA Origins
grant NAG5-13050. KMP is supported by NASA through Hubble Fellowship
grant \#01201.01 awarded by the Space Telescope Science Institute,
which is operated by the Association of Universities for Research in
Astronomy, Inc., for NASA, under contract NAS 5-26555. We thank the
Lorentz Center in Leiden for hosting several meetings that contributed
to this paper. This publication makes use of data products from the
Two Micron All Sky Survey, which is a joint project of the University
of Massachusetts and the Infrared Processing and Analysis
Center/California Institute of Technology, funded by the National
Aeronautics and Space Administration and the National Science
Foundation.

%

\end{document}